\documentclass[prl,aps,a4,epsfig]{revtex4}
\usepackage{amscd}
\usepackage{epsfig}

\begin{document}
\title{Geometric quantum computation using fictitious spin-1/2 subspaces of strongly dipolar coupled nuclear spins} 
\author{T. Gopinath and Anil Kumar\footnote{Electronic mail: anilnmr@physics.iisc.ernet.in}}
\affiliation{{\small \it NMR Quantum Computing and Quantum Information Group.\\Department of Physics, and NMR Research Centre.\\
Indian Institute of Science, Bangalore - 560012, India.}}

\begin{abstract}
Geometric phases have been used in NMR, to implement controlled
phase shift gates for quantum information processing, only in weakly coupled systems in which the individual spins can be identified as qubits.  
In this work,  we implement controlled phase shift gates in strongly coupled systems, by using 
non-adiabatic geometric phases, obtained by evolving the magnetization of fictitious spin-1/2 subspaces, over a closed loop on the Bloch sphere. The 
dynamical phase accumulated during the evolution of the subspaces, is refocused by a spin echo pulse sequence and  by setting the delay of 
transition selective pulses such that the evolution under the homonuclear coupling makes a complete $2\pi$ rotation. A detailed theoretical 
explanation of non-adiabatic geometric phases in NMR is given, by using single transition operators.
Controlled phase shift gates, two qubit Deutsch-Jozsa algorithm and parity algorithm in a qubit-qutrit system have been implemented in various strongly dipolar 
coupled systems obtained by orienting the molecules in liquid crystal media.
\end{abstract}

\maketitle

\section{I. Introduction}
The concept of using quantum systems for information processing was first introduced by Benioff \cite{beni}. In 1985 Deutsch described quantum computers 
which exploit the superposition of multi particle states, thereby achieving massive parallelism \cite{deu}.  
Researchers have also studied the possibility of solving certain types of problems more efficiently than can be done on conventional computers 
\cite{fey,ben,deu,loy}. 
Quantum computing and quantum information processing, requires the ability to execute conditional dynamics between two qubits \cite{bare}. 
Several techniques are being exploited including nuclear magnetic resonance 
\cite{nmr1,nmr2,nmr3,nmr5,jones1}. The experimental 
realization has been limited by many factors, such as decoherence, difficulty in achieving higher 
number of qubits and controlled manipulation of qubits with higher fidelity \cite{pw,deujoz,gr,gru,db,ic}. A recent development in this field is to use 
geometric phases to 
build fault tolerant controlled phase shift gates \cite{jones}.

Berry's discovery of geometric phase \cite{berry} accompanying cyclic adiabatic evolution has triggered an immense effects in holonomy effects in
quantum mechanics and has led to many generalizations. Simon\cite{simon} explained that this geometric phase could be viewed as a consequence of
parallel transport in a curved space appropriate to the quantum system. 
In nuclear magnetic resonance, geometric phase was first verified by Pines et.al.
in adiabatic regime \cite{pines}. A similar approach was used by Jones, to implement controlled phase shift gates by geometric phases in a two qubit system
formed by a weak J coupling \cite{jones2,jones3}. 
However the adiabatic condition is not satisfied in many realistic cases because of the long operation
time \cite{zhu} . Hence it is difficult to experimentally realize quantum computation with adiabatic evolutions, particularly for systems having short decoherence
time. To overcome this disadvantage, it was proposed to use the non-adiabatic cyclic geometric phase (Aharonov and Anandan phase \cite{ahar}) to construct 
geometric quantum gates \cite{wang}. These gates not only have faster gate operation time, but also have intrinsic features of the geometric phase, and 
hence robust against certain types of operational errors \cite{zhu,wang}. For a non-adiabatic cyclic evolution, the total phase difference between the 
initial 
and final states, consists of both the geometric and dynamical phases. Therefore, to obtain only the non-adiabatic geometric phase, one has to remove the 
dynamical component. Non-adiabatic geometric phase in NMR was also first verified by Pines et.al \cite{pines1}.
Non adiabatic geometric phases in NMR, were used to implement controlled phase shift gates, Deutsch-Jozsa (DJ) and Grover algorithms in
a two qubit
system formed by weak J coupling \cite{rana1}.

 Most of the NMR Quantum information processing (QIP) experiments have utilized systems having indirect spin-spin couplings (scalar J couplings)
\cite{nmr1,nmr2,nmr3,nmr4,nmr5}. Since these couplings are mediated via the covalent bonds, the number of coupled spins and hence the number of qubits 
is limited to a few qubits \cite{jones1}. Another approach is to use direct dipole-dipole couplings between the spins which are larger in magnitude
\cite{fung,mahesh,mahesh1,mar,yan,rana}. In liquids the 
dipolar couplings are averaged to zero due to rapid isotropic reorientations of the molecules and in rigid solids there are too many couplings yielding 
broad lines\cite{ern}. In molecules partially oriented in anisotropic media, like liquid crystals, one obtains partially averaged intra molecular dipolar 
couplings with only a finite number of dipolar 
coupled spins, yielding finite number of sharp NMR resonances \cite{fung}. However in such cases often the dipolar couplings are large or comparable to chemical shifts 
differences between the coupled 
spins, leading to spins becoming strongly coupled. The method of spin selective pulses and J evolution used in NMR-QIP of weakly coupled spins 
(liquid state NMR-QIP), leads to complications in strongly coupled spins because of the non secular terms in the Hamiltonian. Therefore such systems have 
been utilized for quantum information processing, by using transition selective pulses and treating the $2^n$ non-degenerate energy levels as an "n" qubit 
system. Preparation of pseudopure states, one qubit DJ algorithm and quantum gates which do not need control of phases, have been implemented in such 
systems using transition selective pulses \cite{fung,mahesh,mahesh1}. Recently an 8 and a 12 qubit systems have been achieved using dipolar coupled spins 
in molecules oriented in liquid crystals \cite{rana2,khitrin}.

In this paper, we 
implement controlled phase shift gates, two qubit DJ algorithm and the parity algorithm in strongly dipolar coupled spins, by using 
non-adiabatic geometric phases, obtained by evolving 
the magnetization of fictitious spin-1/2 subspaces (formed by pairs of energy levels). This method requires efficient transition selective pulses and 
refocusing 
of Hamiltonian evolution. The physical systems chosen, contain  a single non secular term $\vec{I_{i}}$.$\vec{I_{j}}$ in the homonuclear dipolar Hamiltonian 
$D_{ij} (3I_{zi}I_{zj}-\vec{I_{i}}.\vec{I_{j}})$, which is refocused by setting the delay of a transition selective pulse such that 
the evolution due to the coupling makes a 
complete $2\pi$ rotation. To the best of our knowledge, this is the first implementation of quantum algorithms in 
strongly coupled nuclear spins by using geometric phases. 

In section(II), we outline the non-adiabatic Geometric phases by using both, Bloch sphere approach and single transition operators \cite{woka}. 
In section(III), we describe the implementation of controlled phase shift gates in $^{13}CH_3I$ partially oriented in ZLI-1132 liquid crystal. 
In section(IV) two qubit DJ algorithm is implemented in dipolar coupled equivalent protons of oriented $CH_3CN$. In section (V) parity 
algorithm is implemented in a qubit-qutrit system of $CH_2FCN$ partially oriented in ZLI-1132, and the results are concluded in section(VI).

\section{II. Non-adiabatic Geometric phases in NMR}

In order to describe the non-adiabatic geometric phases in NMR, we discuss the evolution of the magnetization of single quantum transitions, under 
transition selective pulses, using 
the approach of single transition operator algebra \cite{woka}, briefly outlined below.

Consider a 4 level system (fig.1a), consisting of 6 pairs (r,s) of energy levels (r=1,2,3, and r $<$ s $\leq$ 4 ). Each pair of energy levels can be considered 
as belonging to a 
fictitious  spin-1/2 subspace (virtual two level system), and the dynamics of the magnetization of the system can be explained by using single transition operators  \cite{woka}. Here 
we assume that the single quantum magnetization of the subspaces (1,2), (2,3) and (3,4), is directly observable, hence one can easily manipulate those 
subspaces. The angular momentum operators 
(single transition Cartesian operators) $I^{(r,s)}_\alpha$ ($\alpha$=x,y,z) have been defined as \cite{woka},

\begin{eqnarray}
\langle i \vert I^{(r,s)}_x \vert j \rangle=\frac {1}{2}(\delta _{ir}\delta _{js} + \delta _{is}\delta _{jr}),
\nonumber\\ 
\langle i \vert I^{(r,s)}_y \vert j \rangle=\frac {i}{2}(-\delta _{ir}\delta _{js} + \delta _{is}\delta _{jr}),
\nonumber\\
\langle i \vert I^{(r,s)}_z \vert j \rangle=\frac {1}{2}(\delta _{ir}\delta _{jr} - \delta _{is}\delta _{js}),  \label{fs}
\end{eqnarray}
where i, j = 1, 2, 3 and 4.
From this definition it follows that \cite{woka},
\begin{eqnarray}
I_x^{(r,s)}=I_x^{(s,r)},\hspace{0.3cm} I_y^{(r,s)}=-I_y^{(s,r)} \hspace{0.2cm} and \hspace{0.2cm} I_z^{(r,s)}=-I_z^{(s,r)}.
\end{eqnarray}

In the following, geometric phases are explained by using Bloch sphere approach \cite{jones3}. Later, we describe the commutation relations of 
single transition operators eqn.(\ref{fs}), and show that geometric phases can be obtained by using pair of transition selective $\pi$ pulses, with 
suitable phases in the xy plane. The 
removal of dynamical phase 
(phase acquired due to internal Hamiltonian evolution) is discussed in sections III, IV and V.

\begin{center}
\begin{figure}
\epsfig{file=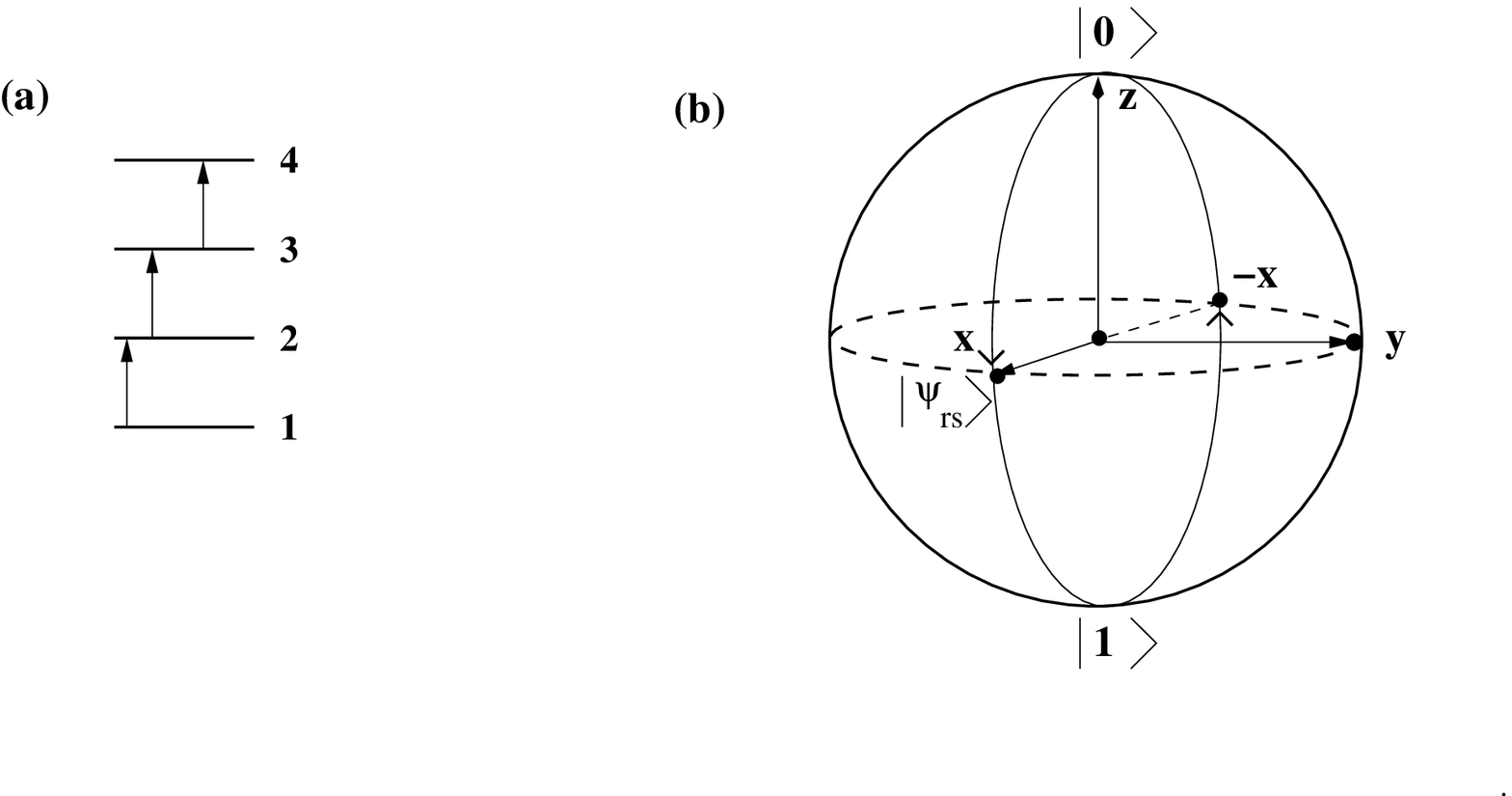,width=18cm}
\caption{} \label{4levels}
\end{figure}
\end{center}

From equation (\ref{fs}), one can see that, the angular momentum operators of (r,s) subspace,  after removing rows and columns containing zeroes, are Pauli spin matrices and hence 
the basis states of (r,s) subspace can be considered as $\vert 0 \rangle$ and  
$\vert 1 \rangle$, which  
can be represented on a Bloch sphere fig.(\ref{4levels}b). In this paper we follow the convention \cite{ic} that the lower state 'r', in the (r,s) subspace, represents 
$\vert 0 \rangle$ and the upper state 's' represents $\vert 1 \rangle$. In general, every two level sub-system at equilibrium, can be considered to be in a 
pseudopure state \cite{cor,ger}. Hence the (r,s) subsystem can be considered as in the pseudopure state $\vert 0 \rangle$. In such 
systems, to encode the phase information, we create initial states, consisting of 
superposition states, in each of the subspaces (1,2), (2,3) and (3,4), given by,

\begin{eqnarray}
\vert \psi_{rs} \rangle =\frac{1}{\sqrt2}(\vert 0 \rangle + \vert 1 \rangle)_{rs}, \hspace{0.5cm} where \hspace{0.3cm}r=1,2,3 \hspace{0.2cm}and\hspace{0.2cm} 
s=r+1. \label{si}
\end{eqnarray}

The state of (r,s) subspace, $\vert \psi_{rs} \rangle$ (eqn.\ref{si}), consist of coherences $I^{(r,s)}_x$ (r=1,2 and 3, s=r+1 ),
which can be
created experimentally by
applying a $(\pi/2)_y$ pulse on the equilibrium state of the whole system. The
coherence $I^{(r,s)}_x$ can in general be written as,

\begin{eqnarray}
I_x^{(r,s)}=\vert \psi_{rs} \rangle \langle \psi_{rs} \vert - \frac{1}{2}I^{(r,s)}, \label{coherences}
\end{eqnarray}

where $I^{(r,s)}$ is the identity matrix of (r,s) subspace and $\vert \psi_{rs} \rangle \langle \psi_{rs} \vert $ represents the density matrix
$\sigma_{(r,s)}$. Since the identity matrix does not contribute to NMR signal,
the density matrix  $\sigma_{(r,s)}=\vert \psi_{rs} \rangle \langle \psi_{rs} \vert$ represents $I_x^{(r,s)}$ and vice versa.

The state $\vert \psi_{rs} \rangle$, can be manipulated, such that it  
goes through a closed loop on the Bloch sphere, thereby acquiring a geometric phase (fig.\ref{4levels}b). The geometric phase acquired 
is respectively $e^{i\Omega/2}$ and $e^{-i\Omega/2}$ for the states $\vert 0 \rangle$ and  $\vert 1 \rangle$, where $\Omega$ is the solid angle 
subtended by the closed loop at the center of the Bloch sphere (fig.\ref{4levels}b) \cite{jones3,pines1}. The unitary operator for the above operation 
is 
a phase shift gate, given for example, for the (2,3) subspace as, 

\begin{eqnarray}
U(2,3)=\pmatrix{ 1&0&0&0 \cr 0&e^{i\Omega/2}&0&0 \cr 0&0&e^{-i\Omega/2}&0 \cr 0&0&0&1 }. \label{U(2,3)}
\end{eqnarray}

The state $\vert \psi_{rs} \rangle$, under the above unitary operator U(2,3), is transformed to $\vert \psi'_{rs} \rangle$, given by,

\begin{eqnarray}
\vert \psi'_{12} \rangle=\frac{1}{\sqrt2}(\vert 0 \rangle + e^{i\Omega/2}\vert 1 \rangle)_{12}
\nonumber\\
\vert \psi'_{23} \rangle=\frac{1}{\sqrt2}( e^{i\Omega/2}  \vert 0 \rangle + e^{-i\Omega/2}\vert 1 \rangle)_{23}
\nonumber\\
\vert \psi'_{34} \rangle=\frac{1}{\sqrt2}( e^{-i\Omega/2} \vert 0 \rangle + \vert 1 \rangle)_{34}   \label{siprime}.
\end{eqnarray}

The states  $\vert \psi'_{rs} \rangle$, in eqn.\ref{siprime}, can be written in the density matrix form $[\sigma'_{rs}=\vert \psi'_{rs} \rangle \langle \psi'_{rs} \vert]$, as,

\begin{eqnarray}
\sigma'_{12}= \frac{1}{2}\pmatrix{ 1&e^{-i\Omega/2} \cr e^{i\Omega/2}&1 \cr},\hspace{0.2cm} 
\sigma'_{23}= \frac{1}{2}\pmatrix{ 1&e^{i\Omega} \cr e^{-i\Omega}&1 \cr}, \hspace{0.2cm}and\hspace{0.2cm}
\sigma'_{34}= \frac{1}{2}\pmatrix{ 1&e^{-i\Omega/2} \cr e^{i\Omega/2}&1 \cr}. \label{sigmaprime}
\end{eqnarray}

In the following we show that, the geometric phase gate U(2,3) can be obtained by, cyclic evolution of $I_x^{(2,3)}$ (state $\vert \psi_{23} \rangle$), 
by means of selective $\pi$ pulses on transition (2,3).

The effect of transition selective pulses can be explained by using the commutators of single transition operators given in eqn.(\ref{fs}).   
If the coherence and the r.f. pulse involves the same transition, then the 
transformation follows the usual commutation relation \cite{woka}, 
\begin{eqnarray}
[I^{(r,s)}_{\alpha},I^{(r,s)}_{\beta}]=i I^{(r,s)}_{\gamma}, \label{com1}
\end{eqnarray}
where $(\alpha,\beta,\gamma)$ is a cyclic permutation of (x,y,z).

The commutator in eqn.\ref{com1}, implies the property that \cite{ern}, 
\begin{eqnarray}
e^{-i\theta I^{(r,s)}_{\gamma}} I^{(r,s)}_{\alpha} e^{i\theta I^{(r,s)}_{\gamma}}=
I^{(r,s)}_{\alpha}cos(\theta)+I^{(r,s)}_{\beta}sin(\theta), \label{pulse}
\end{eqnarray}

which can be interpreted as, rotation of $I^{(r,s)}_{\alpha}$ towards $I^{(r,s)}_{\beta}$, around the $\gamma$ axis of (r,s) subspace.

If the pulse is applied on the transition (r,s), and a connected transition (t,r) is observed, then the commutation relations become \cite{woka},
\begin{eqnarray}
[I^{(t,r)}_x,I^{(t,s)}_x]=\frac{i}{2}I^{(r,s)}_y  \nonumber \\ and \hspace{0.5cm}
 [I^{(t,r)}_y,I^{(t,s)}_x]=\frac{i}{2}I^{(r,s)}_x. \label{com2}
\end{eqnarray}

The commutator relations of eqn.(10) imply that:
\begin{eqnarray}
e^{-i\theta I^{(r,s)}_{y}} I^{(t,r)}_{x} e^{i\theta I^{(r,s)}_{y}}=
I^{(t,r)}_{x}cos(\frac{\theta}{2})+I^{(t,s)}_{x}sin(\frac{\theta}{2}) \nonumber 
 \\ and \hspace{0.5cm} 
e^{-i\theta I^{(r,s)}_{x}} I^{(t,r)}_{y} e^{i\theta I^{(r,s)}_{x}}=
I^{(t,r)}_{y}cos(\frac{\theta}{2})+I^{(t,s)}_{x}sin(\frac{\theta}{2}). 
\end{eqnarray}

\begin{center}
\begin{figure}
\epsfig{file=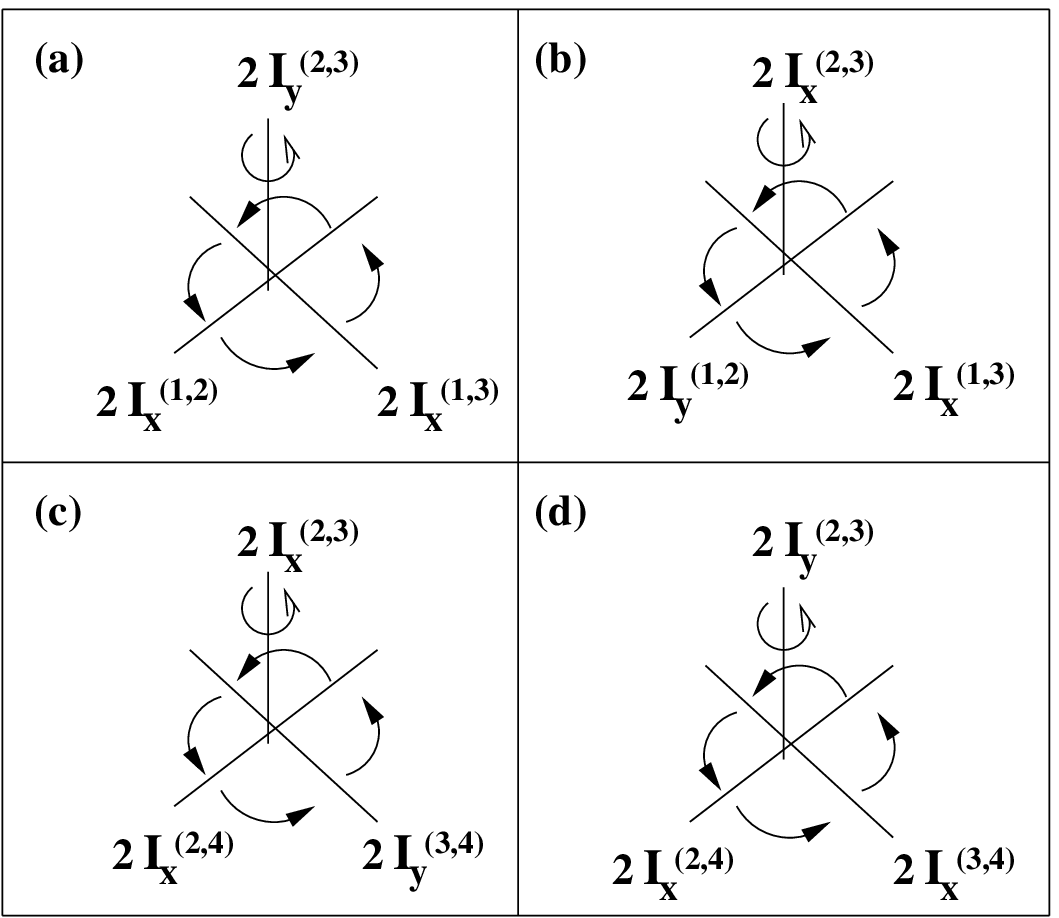,width=10cm}
\caption{} \label{subspaces}
\end{figure}
\end{center}

The first commutator in eqn.(\ref{com2}), implies that, the coherence $I^{(t,r)}_x$ is rotated towards $I^{(t,s)}_x$, when the pulse is applied on transition (r,s), 
about 
y-axis. Whereas the second commutator in eqn.(\ref{com2}), implies that, $I^{(t,r)}_y$ is rotated towards $I^{(t,s)}_x$, under the r.f pulse, 
applied on transition (r,s) about the x-axis. 
Unlike in eqn.(\ref{com1}), the factor (1/2) in eqn.(\ref{com2}), indicates that, the angle of rotation is halved, when the transformed operator and 
the rotation operator have only one 
index in common. Furthermore the commutation relations of eqn.(\ref{com2}) are not cyclic. The commutation rules, given in eqn.(\ref{com2}), can also be 
visualized in three dimensional subspaces spanned by three orthogonal operators each, as shown in fig.\ref{subspaces}, for the subspaces (1,2) and (3,4), when the pulse is applied on (2,3) 
subspace. It may be noted that, in certain rotations the single quantum magnetization will evolve into a 
double or zero quantum coherence.

 We now apply two consecutive selective 
$\pi$ pulses 
on coherence $I_x^{(2,3)}$ (or state $\vert \psi_{23} \rangle$, eqn(\ref{si}) and fig.(\ref{4levels}b)), with phases respectively given by, 
$y$ and $(y+\pi+\phi)$ in the xy plane. The evolution of the coherences $I_x^{(2,3)}$, $I_x^{(1,2)}$ and $I_x^{(3,4)}$  
can be calculated by using the commutation relations given in eqn.s (\ref{com1}) and 
(\ref{com2}) (fig.\ref{subspaces}),      

\begin{eqnarray}
\begin{CD}
I^{(2,3)}_x @>I_y^{(2,3)}(\pi)>> -I^{(2,3)}_x @>I_{(y+\pi+\phi)}^{(2,3)}(\pi)>> cos(2\phi) I^{(2,3)}_x -sin(2\phi) I^{(2,3)}_y,\\
I^{(1,2)}_x @>I_y^{(2,3)}(\pi)>> I^{(1,3)}_x @>I_{(y+\pi+\phi)}^{(2,3)}(\pi)>> cos\phi I^{(1,2)}_x +sin\phi I^{(1,2)}_y,\\  
I^{(3,4)}_x @>I_y^{(2,3)}(\pi)>> -I^{(2,4)}_x @>I_{(y+\pi+\phi)}^{(2,3)}(\pi)>> cos\phi I^{(3,4)}_x +sin\phi I^{(3,4)}_y.
\end{CD}
\end{eqnarray}

The unitary operator associated with above operation (eqn.12) is identical to the phase shift gate U(2,3) of eqn.\ref{U(2,3)}, with $\Omega/2=\phi$. 
In other words, the transformed coherences of equation (12), are identical to $(\sigma'_{rs}-(1/2)I^{(r,s)})$ of eqn.\ref{sigmaprime}, 
with $\phi=\Omega/2$. Hence one can conclude that, two selective $\pi$ pulses, having phases $y$ 
and $(y+\pi+\phi)$ respectively, applied on coherence $I_{(x)}^{(2,3)}$, evolves the state $\vert \psi_{23} \rangle$, over a closed loop on the Bloch sphere, 
and the solid angle subtended by the closed loop, at the center of the sphere is $\Omega=2\phi$. However, in case of a $\pi$-phase shifted gate 
($\phi=\pi$), 
the two $\pi$ pulses, $(\pi)_y$ and $(\pi)_{(y+\pi+\pi=y)}$, shown in fig.(\ref{4levels}b), can be combined in to a single $(2\pi)_y$ 
pulse. Such pulses are used in sections 
III, IV and V.

It is to be noted that, if the coherence of the (2,3) subspace is $I^{(2,3)}_{(\theta'+x)}$ in the xy plane, then the phases of $\pi$ pulses will be 
$(y+\theta')$ and $(y+\theta'+\pi+\phi)$, and a $\pi$-phase shifted gate is achieved by applying $(2\pi)_{(y+\theta')}$ pulse. And if the (2,3) subspace 
is at equilibrium state, $I_{z}^{(2,3)}$, the phases of $\pi$ pulses will be $(\theta'')$ and $(\theta''+\pi+\phi)$, where $\theta''$ can be any axis in the 
xy plane, and a $\pi$-phase shifted gate is achieved by applying $(2\pi)_{\theta''}$ pulse.

From the above discussion it is concluded that, cyclic evolution of a two level subspace, by means of transition selective $\pi$ pulses, acquires a 
geometric phase which also effects the other connected subspaces.

 \section{III. Experimental Implementation of the controlled phase shift gate}
The system chosen is, dipolar coupled carbon and protons of the methyl group of $^{13}CH_3I$, partially oriented in ZLI-1132 liquid crystal. The three 
protons are, chemically and magnetically equivalent, and the total Hamiltonian is given by, 

\begin{equation}
H=\omega_{C} I_{z}+ 3\omega_{H} S_{z}+ \sum_{i,j=1,2,3,\;i<j,\;}2\pi[(J_{CH}I_zS_{zi}
+2D_{CH}I_{z}S_{z_i}+D_{HH}(3S_{z_i}S_{z_j}-S_i.S_j)], \label{hamiltonian1}
\end{equation}

where $\omega_{H}$ and $\omega_{C}$ are chemical shifts of protons and carbon respectively, $D_{HH}$ is the residual 
dipolar coupling between identical protons (homonuclear coupling), and $J_{CH}$ and $D_{CH}$ are couplings of carbon (I) with protons (S). 
The scalar J coupling between identical protons does not effect the spectra in this case \cite{ern}.

\begin{center}                                                   
\begin{figure}                                                                  
\epsfig{file=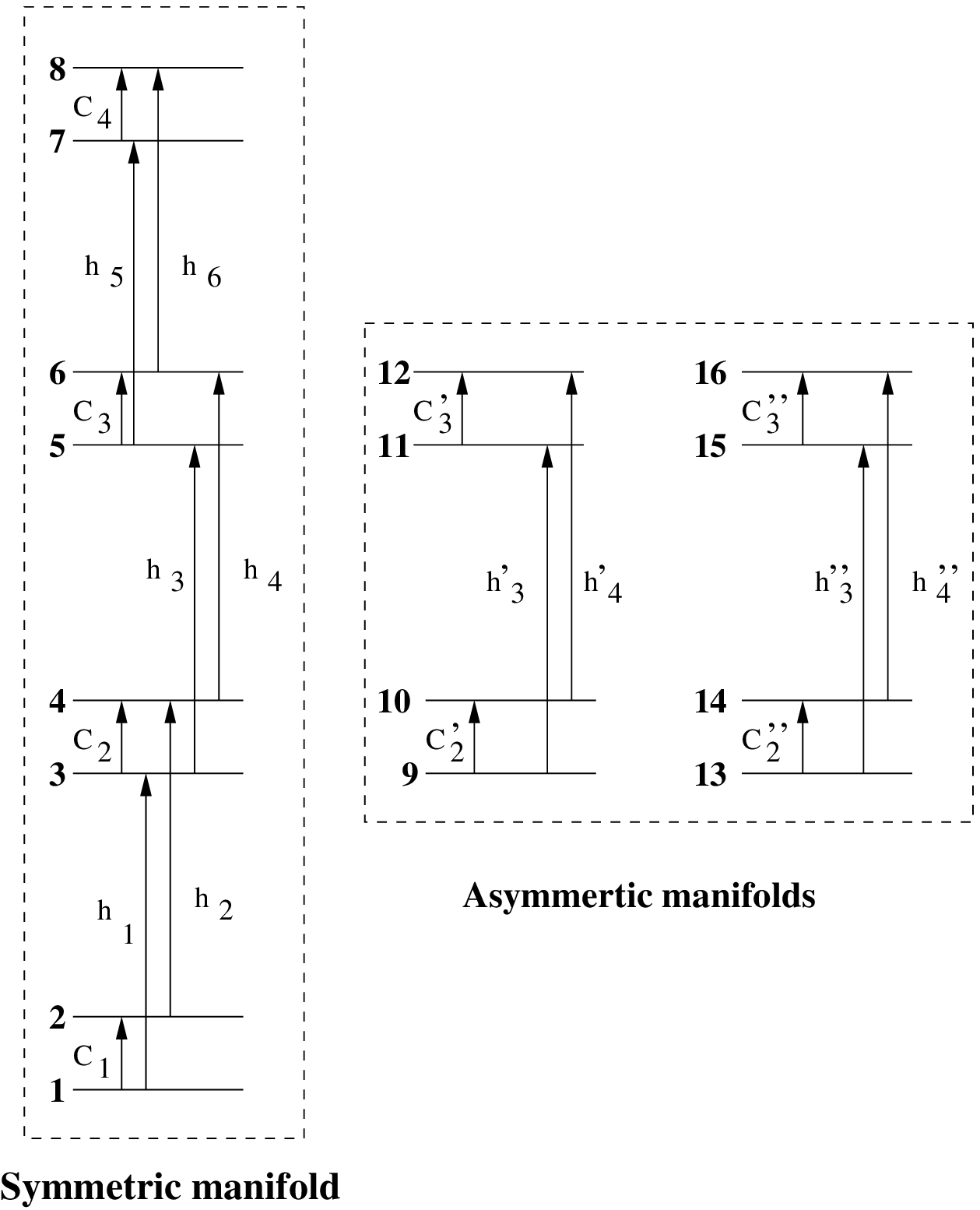,width=12cm}                          
\caption{}    \label{energy levels of 13CH3CN}                                                                  
\end{figure}                                                                    
\end{center}

The energy level 
diagram (fig.\ref{energy levels of 13CH3CN}) of this system consists of 16 energy levels, eight of which belong to a symmetric manifold, and the other 
eight to two asymmetric manifolds \cite{fung}. The eight energy levels of the symmetric manifold can be considered 
as a 3-qubit system. 
In this system recently pair of pseudo pure states (POPS) and quantum gates (other than controlled phase shift gates) have been 
implemented by applying transition selective pulses \cite{mahesh,fung}. 
In this work we implement controlled phase shift gates by using non-adiabatic geometric phases obtained by using transition selective pulses.

\begin{center}
\begin{figure}
\epsfig{file=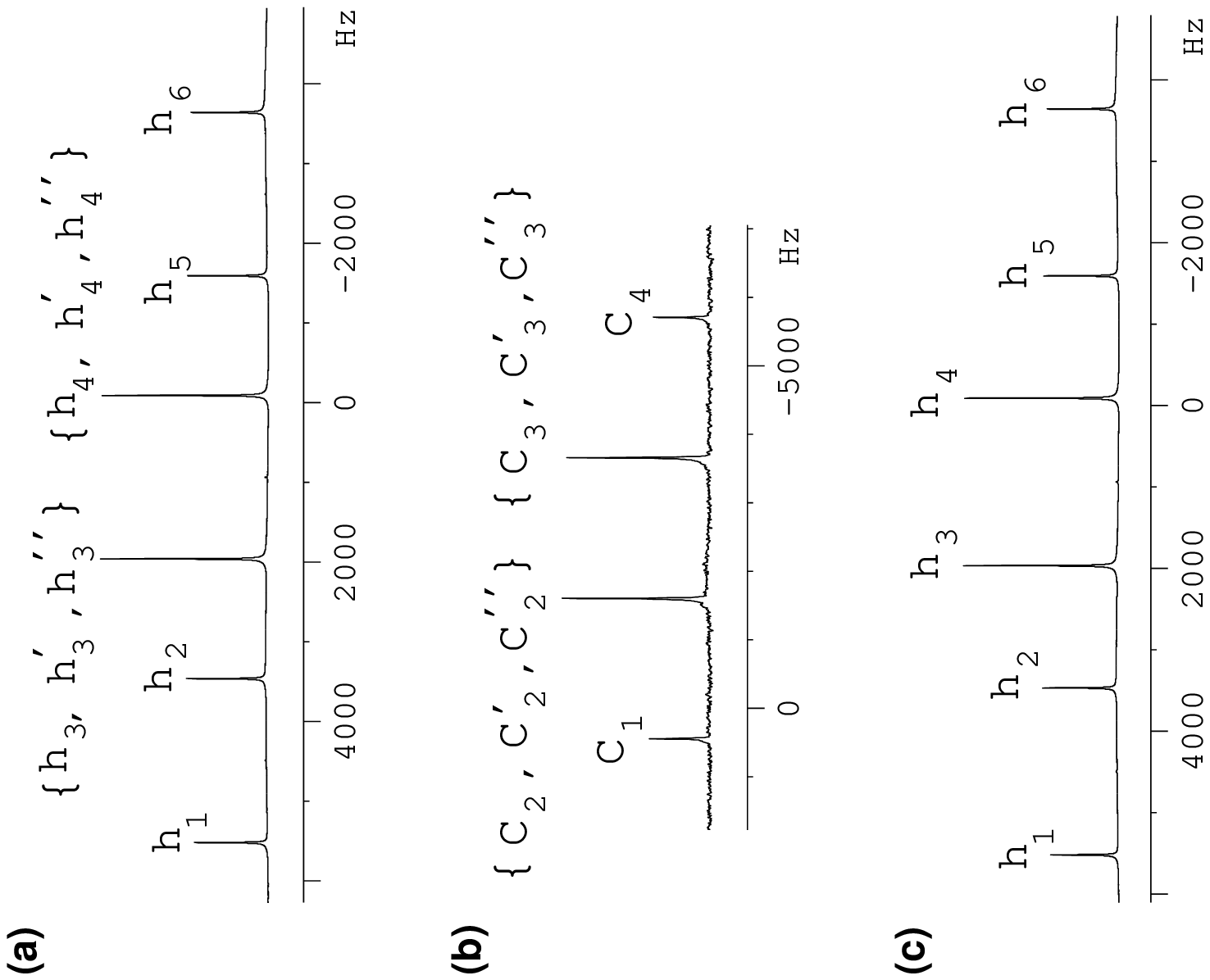,width=12cm,angle=270}
\caption{} \label{equilibrium spectra of 13CH3CN}
\end{figure} 
\end{center}

The equilibrium $^{1}H$ and  $^{13}C$ spectra are given in 
fig.(\ref{equilibrium spectra of 13CH3CN}a) and fig.(\ref{equilibrium spectra of 13CH3CN}b) respectively, wherein the various 
transitions are 
labeled in accordance with the notation used in fig.\ref{energy levels of 13CH3CN}. 
Since several transitions of symmetric and asymmetric manifolds, are degenerate, selection of symmetric states (SOSS) is performed to select the magnetization 
of the symmetric manifold \cite{mahesh}. However, since in the present experiments only the proton transitions are observed, we perform only a 
partial SOSS, in which the proton magnetization of symmetric manifold, is selectively retained, as explained below. 

\begin{center}
\begin{figure}
\epsfig{file=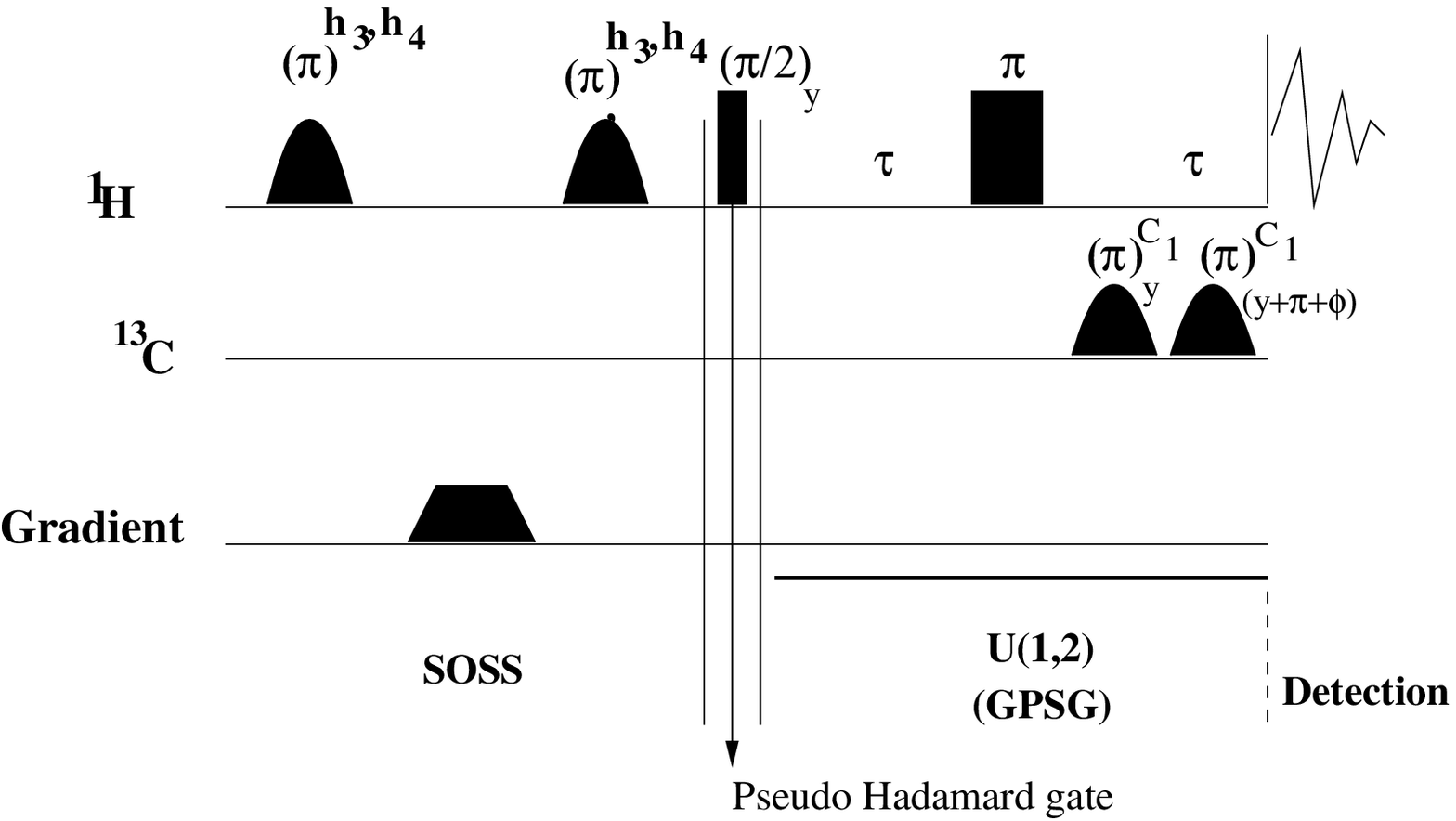,width=12cm}
\caption{}   \label{pulse sequence for 13CH3CN}
\end{figure}
\end{center}

The entire experiment consists of 4 steps (as shown in fig.\ref{pulse sequence for 13CH3CN}).

(i) Selection of symmetric states (SOSS): 
 The proton magnetization from asymmetric manifolds is saturated by using the pulse sequence 
$(\pi) ^{{h_3},{h_4}}$-$G_z$-$(\pi) ^{{h_3},{h_4}}$ (fig.\ref{pulse sequence for 13CH3CN}) , where $(\pi) ^{{h_3},{h_4}}$ represents two selective $\pi$ pulses on 
$h_3$ and $h_4$ 
transitions, and $G_z$ is the pulsed field gradient \cite{mahesh}. The $\pi$ pulse on transitions $h_3$ and $h_4$ acts as a $\pi/2$ pulse on $h'_3$ and 
$h'_4$ \cite{mayne}, and a gradient dephases the coherences $h'_3$ and $h'_4$. The second $\pi$ pulse brings back the magnetization of symmetric 
manifold 
to its equilibrium value. Proton spectrum obtained by a small angle pulse followed by SOSS  (fig.\ref{equilibrium spectra of 13CH3CN}c), has the 
intensities in the expected ratio 3:3:4:4:3:3, confirms the retention 
of proton magnetization only from the symmetric manifold. 

(ii) An on resonance $(\pi/2)_y$ pulse is applied on the proton channel, which creates the proton coherences, $h_1$, $h_2$, $h_3$, $h_4$, $h_5$ and $h_6$ . 
These coherences does not evolve under the chemical shift, since pulse is applied at resonance.  

(iii) Controlled phase shift gates: As described in the previous section, cyclic evolution of a two level subspace by means of transition selective $\pi$  
pulses, acquire a geometric phase. 
In order to get
 only the geometric phase, one has to refocus the dynamical phase acquired due to internal Hamiltonian evolution. Spin echo sequence ($\tau - \pi - \tau$) is 
applied on protons 
(fig.\ref{pulse sequence for 13CH3CN}) to refocus hetero nuclear ($^{13}C$-$^{1}H$) coupling (dipolar and J). The geometric phase gates 
are obtained by applying two selective $\pi$ pulses on carbon transitions, (fig.\ref{pulse sequence for 13CH3CN}), and the duration of the pulses is 
chosen such that the evolution of proton coherences under the homonuclear coupling makes a complete rotation (in multiple of $2\pi$).

\begin{center}
\begin{figure}
\epsfig{file=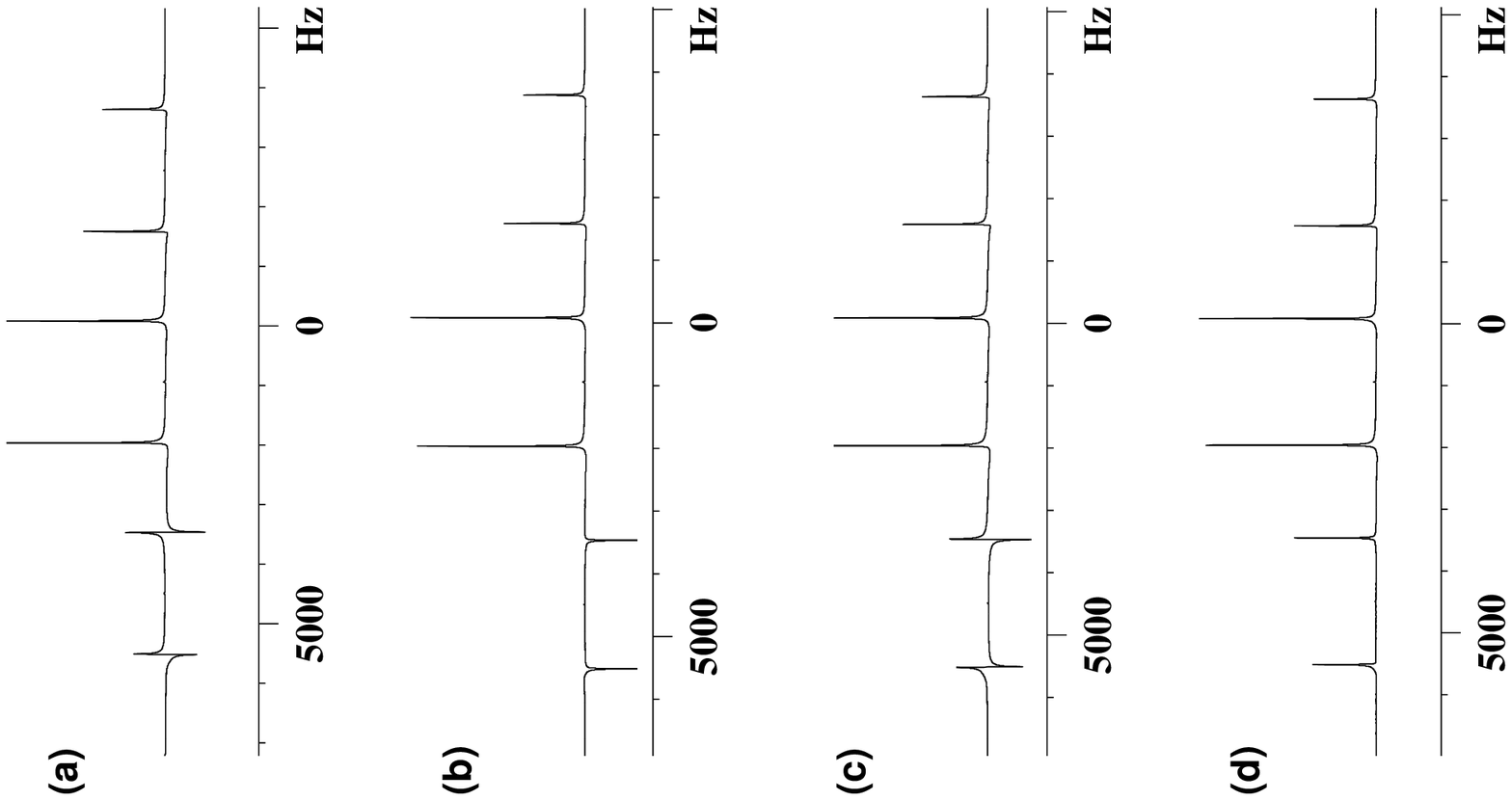,width=12cm,angle=270}
\caption{}   \label{cphaseshift gates1}
\end{figure}
\end{center}

\begin{center}
\begin{figure}
\epsfig{file=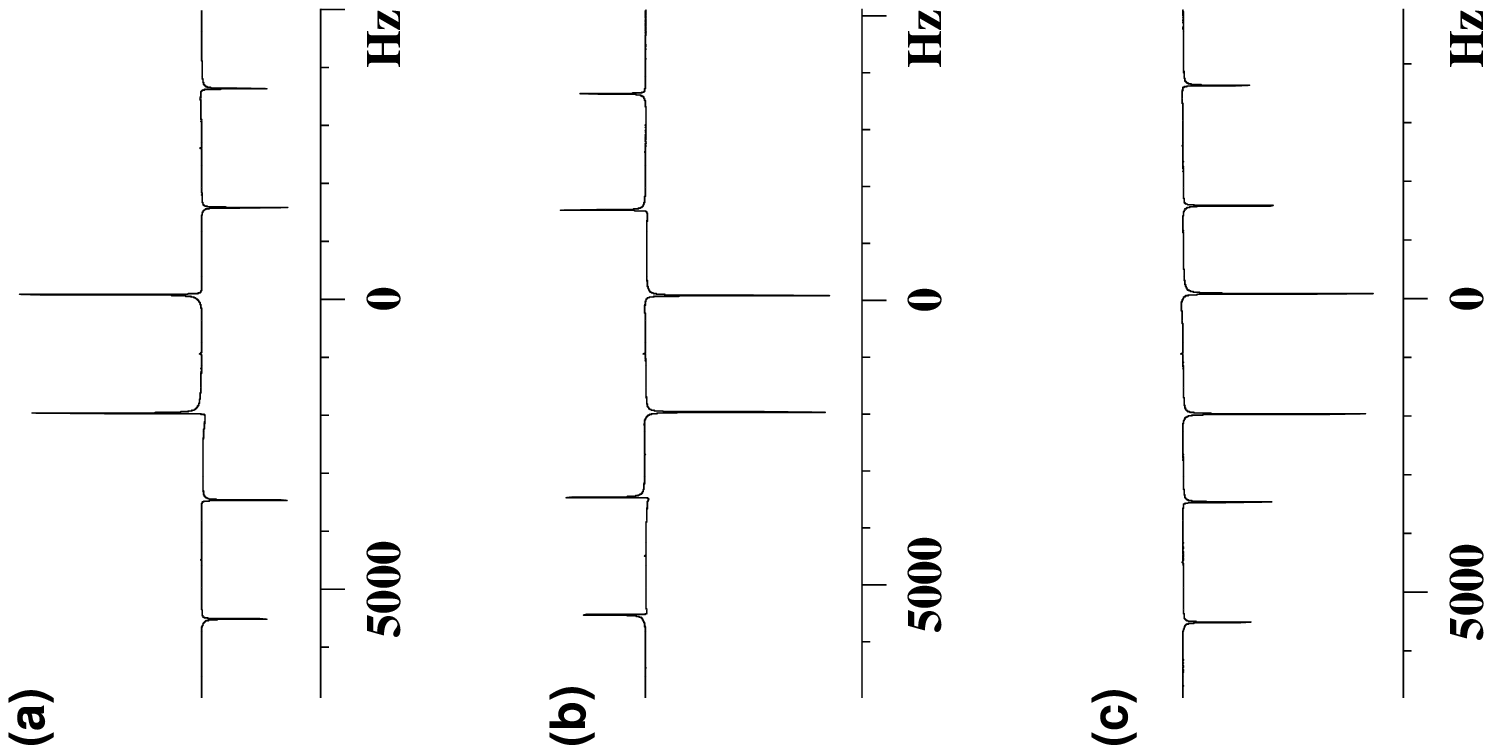,width=12cm,angle=270}
\caption{}   \label{cphaseshift gates2}
\end{figure}
\end{center}

(iv) Detection: The phases of proton coherences $h_1$...$h_6$
 (figures \ref{cphaseshift gates1} and \ref{cphaseshift gates2}), are only due to cyclic geometric evolution of subspaces, confirming the implementation 
of phase shift gates, as explained below.

In fig.(\ref{cphaseshift gates1}), geometric phases are obtained in (1,2) subspace , by applying two consecutive phase shifted selective $\pi$ pulses on 
transition  $C_1$. The phase of the first  
pulse is $y$, and that of second pulse is $(y+\pi+\phi)$, where $\phi$=$\pi/2$, $\pi$, $3\pi/2$ and $2\pi$, for fig.s \ref{cphaseshift gates1}(a), 
\ref{cphaseshift gates1}(b), \ref{cphaseshift gates1}(c), and \ref{cphaseshift gates1}(d) 
respectively. The unitary operator obtained by the above operation is, U(1,2)=diag[$e^{i\phi}$,$e^{-i\phi}$,1,1,1,1,1,1 ]. As expected, 
only the proton 
coherences
 $h_1$ and $h_2$ are effected by the unitary operator U(1,2), whereas the other proton coherences remain un affected. For example in 
fig(\ref{cphaseshift gates1}a), the coherences $I_x^{(1,3)}(h_1)$ and $I_x^{(2,4)}(h_2)$, are transformed to $I_y^{(1,3)}$ and $-I_y^{(2,4)}$ giving dispersive 
signals of opposite sign, confirming the unitary operator U(1.2), 
for $\phi=\pi/2$.

Figure(\ref{cphaseshift gates2}) shows the implementation of $(\pi)$-phase shift gates on two unconnected subspaces; 
U[(3,4),(5,6)]=diag[1,1,-1,-1,-1,-1,1,1 ], U[[(1,2),(3,4)]]=diag[-1,-1,-1,-1,1,1,1,1] and 
U[(1,2),(5,6)]=diag[-1,-1,1,1,-1,-1,1,1 ] (fig.s \ref{cphaseshift gates2}a, \ref{cphaseshift gates2}b and \ref{cphaseshift gates2}c respectively). 
Each of these gates require two $(2\pi)_y$ pulses on unconnected subspaces.  
U[(3,4),(5,6)] is implemented by applying $(2\pi)_y$ pulses on transitions $C_2$ and $C_3$, 
U[(1,2),(3,4)] is implemented by applying  
$(2\pi)_y$ pulses on transitions $C_1$ and $C_2$. Whereas U[(1,2),(5,6)] is implemented by applying $(2\pi)_y$ pulses on transitions $C_1$ and $C_3$. The 
signs of various transitions in the observed spectra of fig. \ref{cphaseshift gates2}, confirm the implementation 
of these phase shift gates. For example, the U[(3,4),(5,6)] gate inverts only the coherences $h_1$, $h_2$, $h_5$ and $h_6$.

\section{IV. Implementation of Deutsch-Jozsa (DJ) algorithm}
\begin{center}
\begin{figure}
\epsfig{file=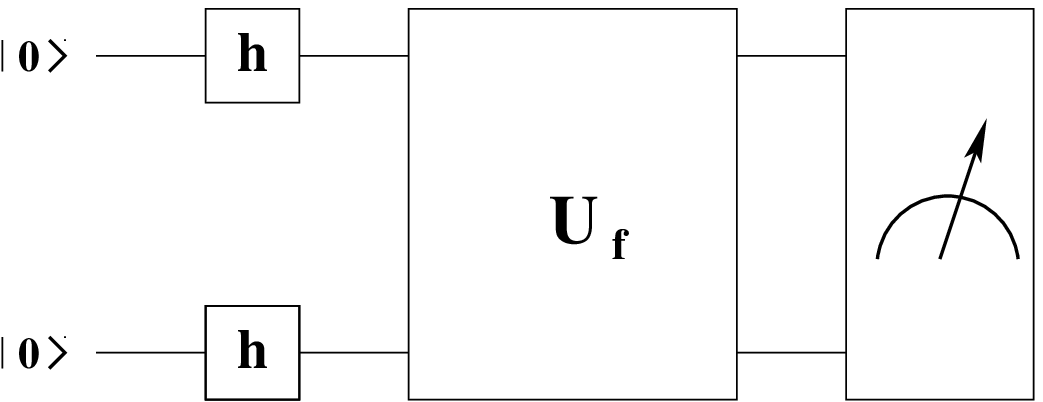,width=10cm}
\caption{}   \label{qcircuit of dj}
\end{figure}
\end{center}

DJ algorithm determines in a single query, whether a given function is constant or balanced \cite{deujoz,cleve}. A function is "constant" if it gives the 
same
output for all inputs, and "balanced" if it gives one output for half the number of inputs and another for the remaining half. Classically for an n bit 
binary function, at least ($2^{n-1}+1$) queries are needed to determine whether the function is constant or
balanced, whereas the DJ algorithm requires only a single query. The Cleve version \cite{cleve} of the DJ algorithm requires an extra qubit
(ancilla qubit) and uses controlled-not gates, whereas Collins version does not require the ancilla qubit but needs controlled phase shift gates \cite{col}. While the Cleve version of DJ algorithm has been implemented both in weakly and strongly coupled systems \cite{kavita,mahesh1}, the Collins version 
has been implemented only in 
weakly coupled systems \cite{mangold}. Here we implement the
Collins version of the DJ algorithm  on a strongly coupled two qubit system, formed by dipolar coupled methyl protons of $CH_3CN$, oriented in ZLI-1132 liquid crystal. For a two qubit system there are 
2 constant and 6 balanced functions, table(1). The quantum circuit of the Collins version of the DJ algorithm is shown in fig.\ref{qcircuit of dj}. The 
algorithm (fig.\ref{qcircuit of dj}) starts with a pure state (pseudo pure state in NMR) $\vert 00 \rangle$ which is converted to a equal superposition
state of the basis states $\vert 00 \rangle$, $\vert 01 \rangle$, $\vert 10 \rangle$, and $\vert 11 \rangle$, by applying a pseudo Hadamard gate on both 
the qubits. Thereafter a unitary operator $U_{f_i}$ (controlled phase shift gate) is applied,  
followed by detection. Theoretically, after
$U_{f_i}$ one
has to apply the Hadamard gate. In NMR, a Hadamard gate is replaced by a pseudo Hadamard gate, which is implemented by a
$(\pi/2)_{y}$ pulse, and for detection another $(\pi/2)_{-y}$ pulse is needed. These two pulses cancel each other and the result of the algorithm is 
available immediately
after $U_{f_i}$. The Unitary operator $U_{f_i}$ can be written as,

\begin{eqnarray}
U_{fi}=\pmatrix{ (-1)^{f_i(00)}&0&0&0 \cr 0&(-1)^{f_i(01)}&0&0 \cr 0&0&(-1)^{f_i(10)}&0 \cr 0&0&0&(-1)^{f_i(11)} }, \label{U_fi}
\end{eqnarray}
 
where i=1 and 2 represent the "constant" function and i=3,4,5,6,7 and 8 represents the "balanced" functions, table 1.

The unitary operators,
$U_{f_1}$, $U_{f_3}$, $U_{f_5}$ and $U_{f_7}$ are identical, up to an overall phase factor, of $e^{i\pi}$, to $U_{f_2}$, $U_{f_4}$, $U_{f_6}$ and $U_{f_8}$ 
respectively. Hence we implement only the unitary operators $U_{f_1}$, $U_{f_3}$, $U_{f_5}$ and $U_{f_7}$.

\begin{center}
\begin{figure}
\epsfig{file=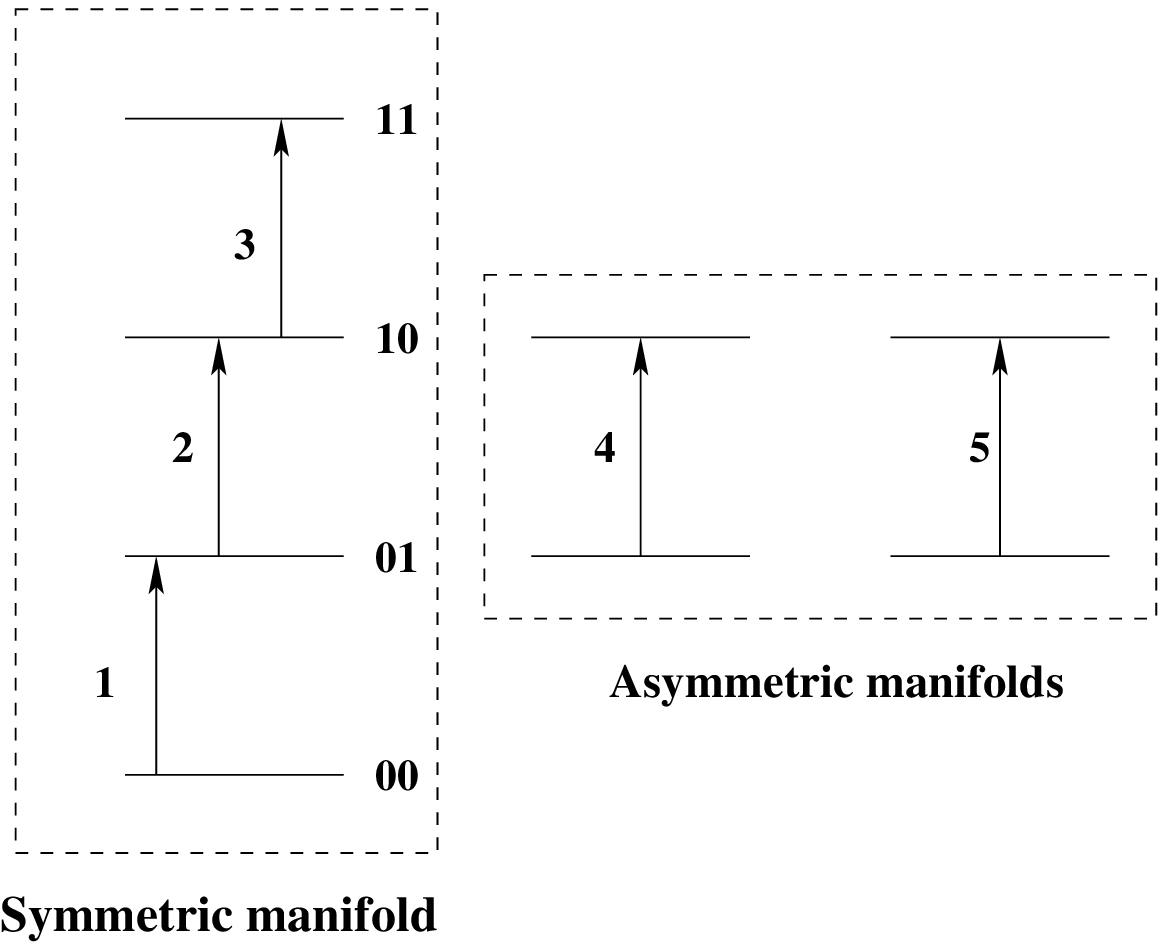,width=12cm}
\caption{}   \label{energy levels of CH3CN}
\end{figure}
\end{center}

\begin{center}
\begin{figure}
\epsfig{file=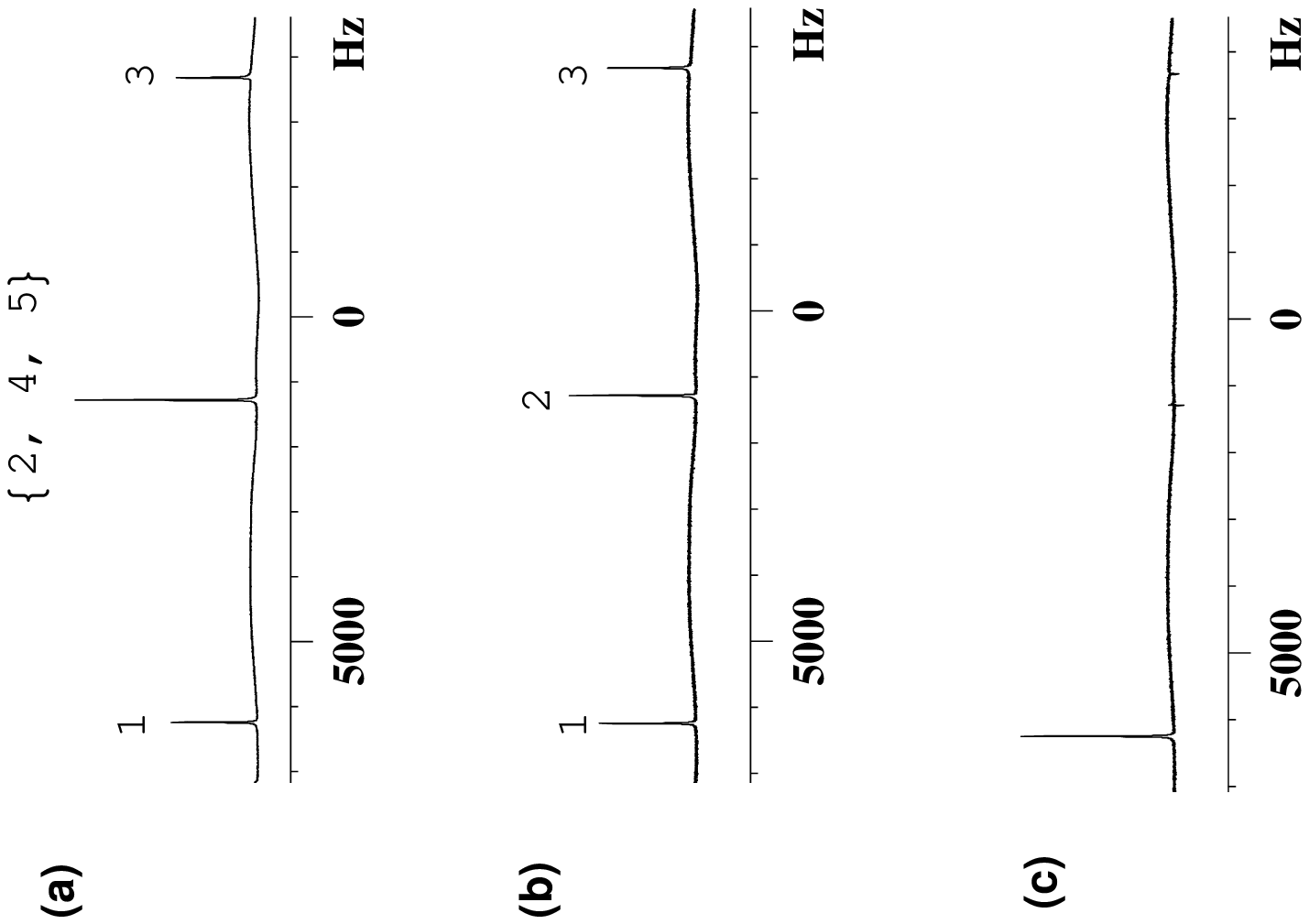,width=12cm,angle=270}
\caption{}   \label{pps of CH3CN}
\end{figure}
\end{center}

The schematic energy level diagram of the dipolar coupled methyl protons of $CH_3CN$ (without C-13 labeling) oriented in ZLI-1132 liquid crystal, 
is given in (fig.\ref{energy levels of CH3CN}). Transitions 
1, 2 and 3 of symmetric manifold 
gives rise to three single quantum transitions at frequencies $\omega_o-3D$, $\omega_o$ and $\omega_o+3D$, with the intensity ratio 3:4:3, 
where D is the partially averaged (residual) dipolar coupling between the protons of $CH_3CN$ and $\omega_o$ is their Larmor frequency. 
Transitions of asymmetric manifold also have 
the frequency $\omega_o$ identical to transition 2. The resultant spectrum is a 1:2:1 triplet as shown in 
fig.\ref{pps of CH3CN}a. The symmetric  
manifold of this system, consisting of four energy levels, can be treated as a two qubit system, and 
the labeling is given in fig.\ref{energy levels of CH3CN}. 

\begin{center}
\begin{figure}
\epsfig{file=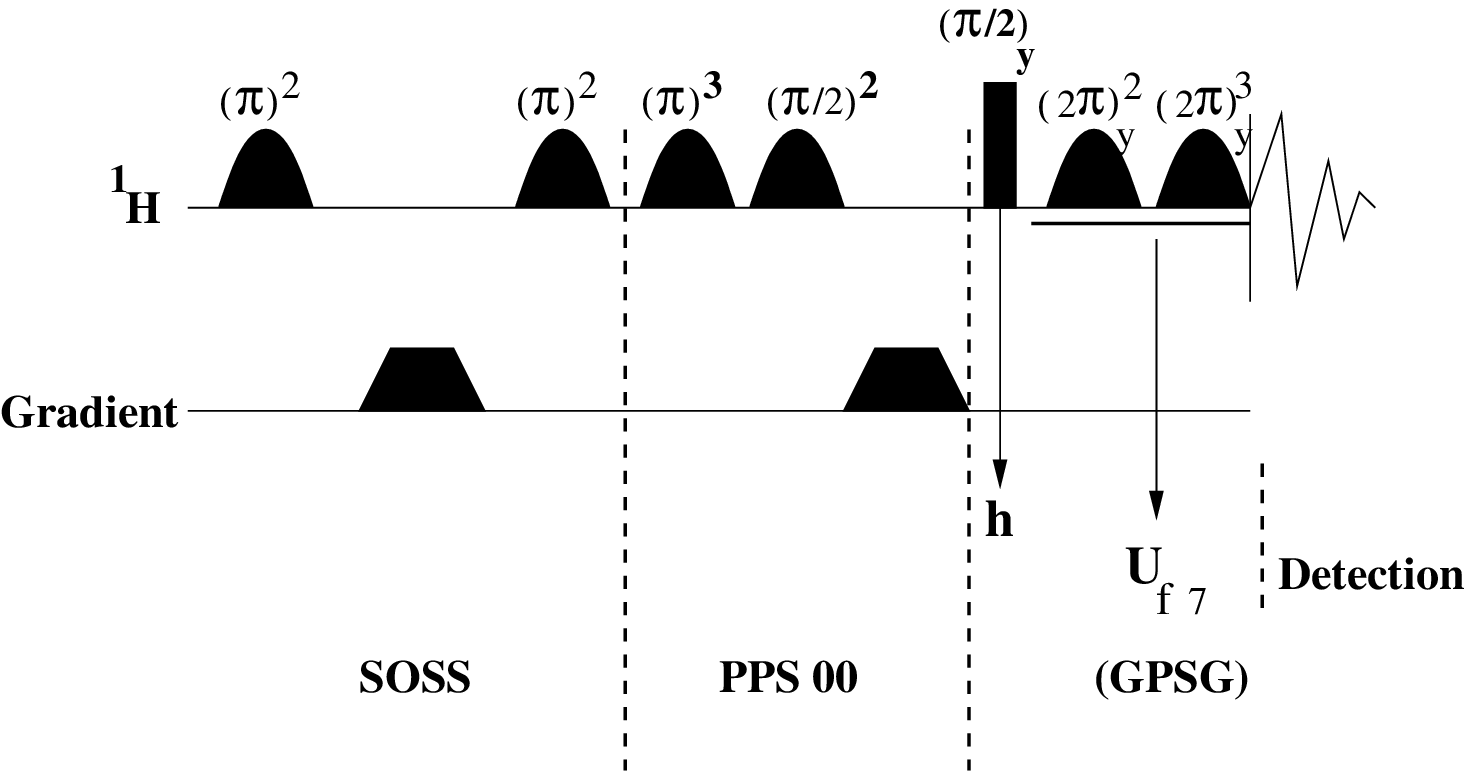,width=12cm}
\caption{}   \label{pulse sequence of CH3CN}
\end{figure}
\end{center}

The total experiment consists of 4 steps (as shown in fig. \ref{pulse sequence of CH3CN}).

(i)Selection of symmetric manifold:
The magnetization of 
asymmetric manifolds which contribute to central transition, is saturated by applying a $\pi$ pulse on transitions 2, which acts as a $\pi/2$ pulse 
on transitions 4 and 5, a gradient pulse dephases the coherences, and the magnetization of the symmetric manifold is retained to its equilibrium value, 
with another 
$\pi$ pulse. The relative integrated intensities of fig.\ref{pps of CH3CN}b (in the ratio 3:4:3) confirm the selection of symmetric manifold. 

(ii)Creation of pseudo pure state (PPS) $\vert00\rangle$ \cite{mahesh}:
The pseudo pure state $\vert00\rangle$ is prepared by applying a $\pi$ pulse on transition 3 (fig. \ref{pulse sequence of CH3CN}) , followed 
by a $\pi/2$ pulse on 
transition 2, a gradient pulse is applied to
daphase the coherences. The proton spectrum of fig.\ref{pps of CH3CN}c, confirm the creation of the $\vert 00 \rangle$ PPS.

(iii) Superposition state followed by $U_{f_i}$: After the creation of $\vert00\rangle$ PPS, a pseudo Hadamard gate ( $(\pi/2)_{y}$ on resonance pulse)
creates a superposition of 4 basis states, and due to on-resonance pulse the coherences do not evolve under the chemical shift. It may be noted that unlike in weakly coupled spins, the state created here is not in uniform superposition 
of eigenstates, since the coefficients of various eigenstates are different \cite{mahesh}. However, as is shown here, the created coherent 
superposition can be 
utilized for quantum parallelism, to distinguish 
different classes of functions. The state of the system after a $(\pi/2)_{y}$ pulse is   
$\vert \psi \rangle=(1/2 \sqrt{2})(\vert 00 \rangle + \sqrt{3} \vert 01 \rangle + \sqrt{3} \vert 10 \rangle + \vert 11 \rangle)$.

The unitary operator $U_{f_i}$ of eqn.(\ref{U_fi}), is 
implemented by using  non-adiabatic geometric phases, obtained by transition selective $(2\pi)_{y}$ pulses. The duration of transition selective pulses (fig. \ref{pulse sequence of CH3CN}) is set such that 
the evolution due to homonuclear $^{1}H-^{1}H$ dipolar coupling, makes a complete $2\pi$ rotation. 
$U_{f_1}$ is a unit matrix which
require no pulses.
$U_{f_3}$ and $U_{f_5}$ are implemented 
by applying a single 
transition selective $(2\pi)_y$ pulse on transitions 2 and 3 respectively. $U_{f_7}$ is implemented by applying two transition 
selective $(2\pi)_y$ pulses on transitions 2 and 3. 
The final state $\vert \psi_{f_i} \rangle$ (=$U_{f_i}\vert \psi \rangle$) can be written as,
\begin{eqnarray}
\vert \psi_{f_1} \rangle=\frac{1}{2\sqrt{2}}(\vert 00 \rangle + \sqrt{3}\vert 01 \rangle + \sqrt{3}\vert 10 \rangle + \vert 11 \rangle  ),
\nonumber\\
\vert \psi_{f_3} \rangle=\frac{1}{2\sqrt{2}}(\vert 00 \rangle - \sqrt{3}\vert 01 \rangle - \sqrt{3}\vert 10 \rangle + \vert 11 \rangle  ),
\nonumber\\
\vert \psi_{f_5} \rangle=\frac{1}{2\sqrt{2}}(\vert 00 \rangle + \sqrt{3}\vert 01 \rangle - \sqrt{3}\vert 10 \rangle - \vert 11 \rangle  ),
\nonumber\\
\vert \psi_{f_7} \rangle=\frac{1}{2\sqrt{2}}(\vert 00 \rangle - \sqrt{3}\vert 01 \rangle + \sqrt{3}\vert 10 \rangle - \vert 11 \rangle  ).
\end{eqnarray}

\begin{center}
\begin{figure}
\epsfig{file=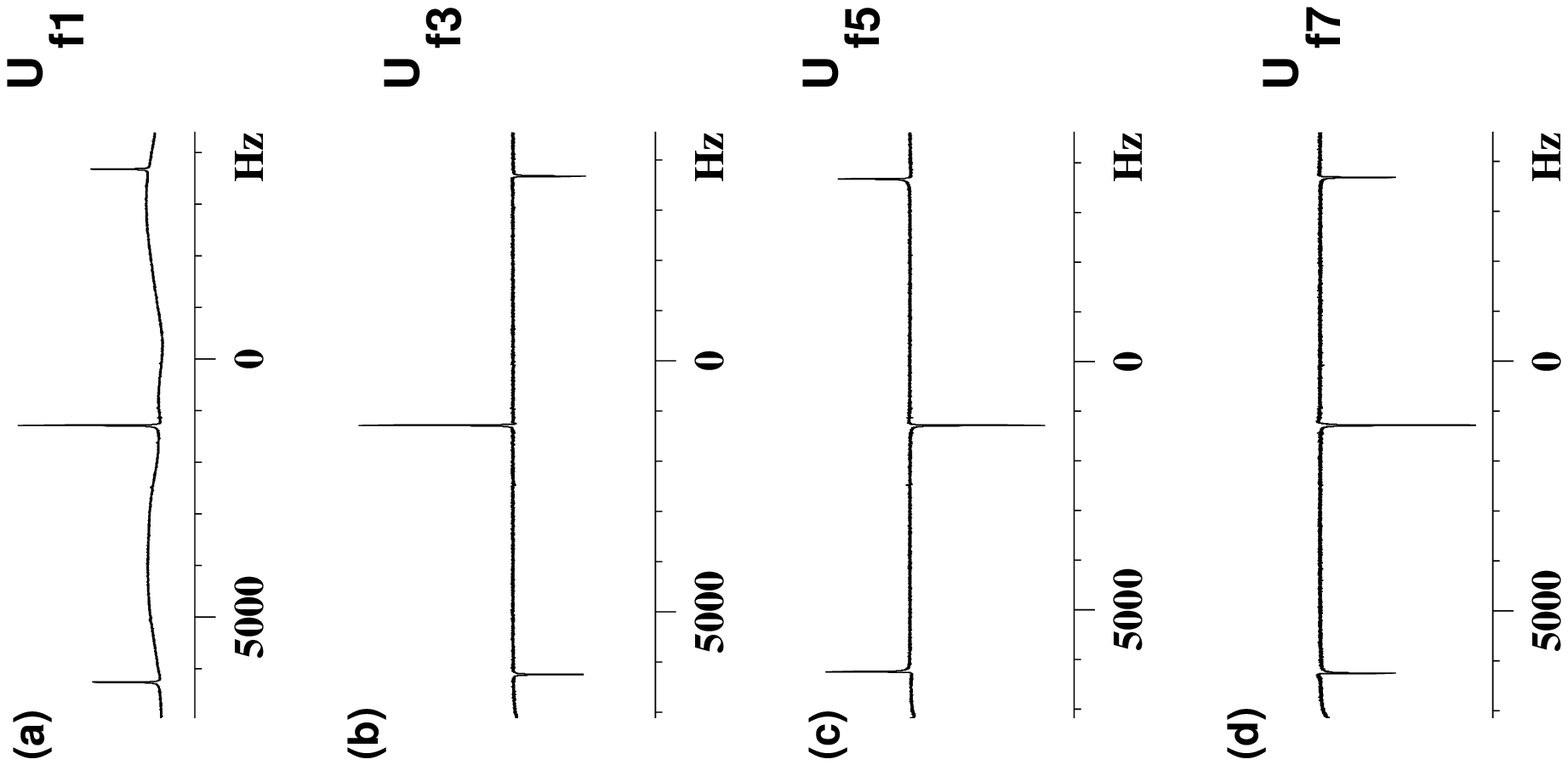,width=12cm,angle=270}
\caption{}   \label{result of dj}
\end{figure}
\end{center}

(iv) Detection: The phases of coherences 1, 2 and 3 confirm the final states $\vert \psi_{f_i} \rangle$. 
For constant function ($\vert \psi_{f_1} \rangle$), none of the peaks are inverted (fig.\ref{result of dj}a), 
whereas for the balanced functions ($\vert \psi_{f_3} \rangle$, $\vert \psi_{f_5} \rangle$ and $\vert \psi_{f_7} \rangle$), atleast one of the peaks is 
inverted (fig.\ref{result of dj}b,\ref{result of dj}c,\ref{result of dj}d respectively). The spectra of figure \ref{result of dj}, confirm the 
implementation of Collins version of two-qbit DJ algorithm in the dipolar coupled spins by using non-adiabatic geometric phase gates.

\section{V. Implementation of Parity algorithm}
The parity of a binary string X=$\{x_1,----,x_N\}$, where $x_i$ is 0 or 1, is even (odd) if x consists of even number (odd number) of 1's. The parity of X 
can be written as a Boolean function P(X)=$x_1 \oplus.....\oplus x_N$ . A classical computer has to call the oracle with each of the N possible 
inputs to determine P(X). 
Beals et. al and Farhi et. al showed that in a quantum computer the minimum number of oracle calls is N/2 \cite{beals,farhi}. The parity algorithm has been 
implemented by Suter et.al \cite{suter} on a weakly coupled two qubit 
system. They have also shown that in case of ensemble quantum computers the complexity is further reduced \cite{suter}. In the quantum version, 
the parity of the binary string X is encoded in the Black box (oracle), such that the black box consists of N boolean variables, 
X=$\{x_1,----,x_N\}$. The task is 
to determine the parity of the binary string, that is encoded in the oracle.

Recently it has been demonstrated that qutrits can be useful for certain purposes of quantum simulation, quantum computations and quantum communication, 
a qutrit has three states $\vert0\rangle$, $\vert1\rangle$ and $\vert2\rangle$ \cite{terhal,flitney,fitzi,rana3,klimov,paul}. 
In this work, we use a quantum circuit similar to the one used in ref.\cite{suter} and 
implement a qubit-qutrit parity algorithm in a single iteration of the oracle, starting with a mixed state. Non-adiabatic geometric phase gates are used to 
implement the required oracles.
The qubit-qutrit system used here is obtained by orienting flouro-acetonitrile in ZLI-1132 liquid crystal. The two identical protons have one 
asymmetric and three symmetric eigen states (labeled as 0, 1, 2 in fig.\ref{energy levels of CH2FCN}). The symmetric manifold of protons can be 
treated as a qutrit and the flourine nucleus as a qubit, yielding a 6 energy level system shown in fig.(\ref{energy levels of CH2FCN}).

\begin{center}
\begin{figure}
\epsfig{file=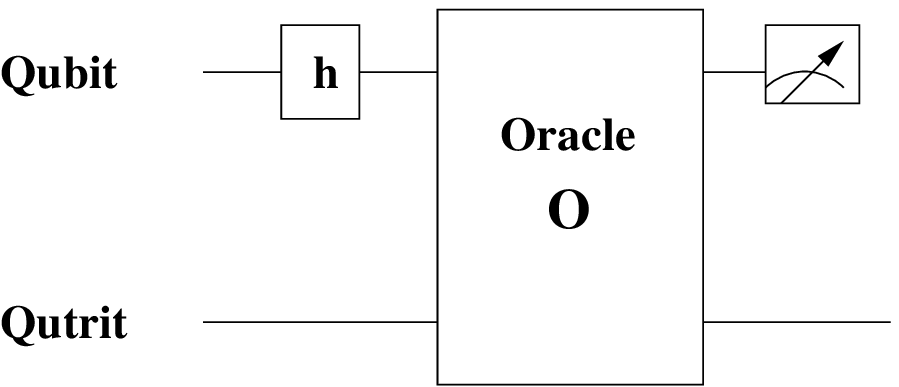,width=10cm}
\caption{}    \label{qcircuit of parity}
\end{figure}
\end{center}

\begin{center}
\begin{figure}
\epsfig{file=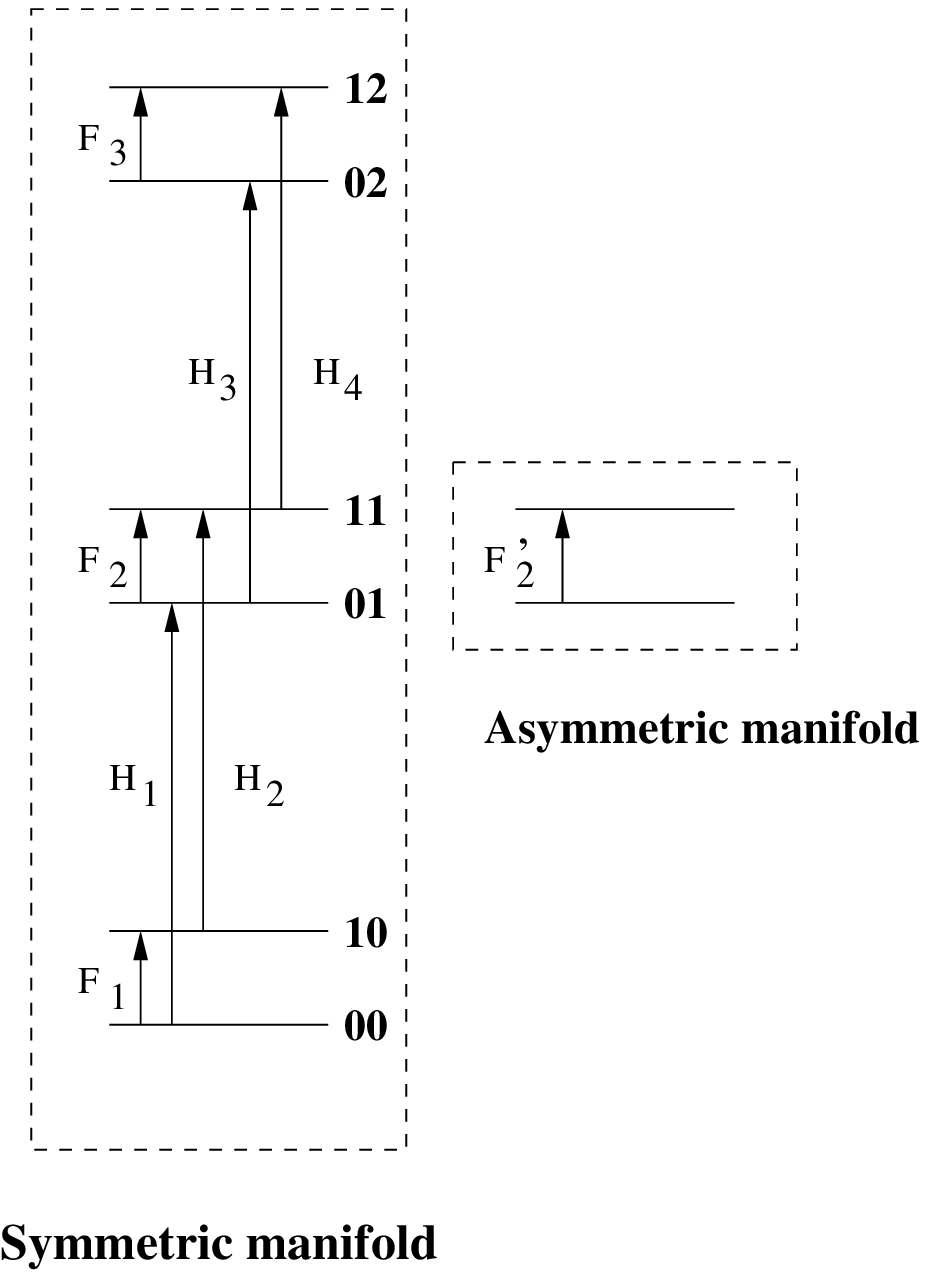,width=8cm}
\caption{}    \label{energy levels of CH2FCN}
\end{figure}
\end{center}

In the quantum circuit of the algorithm, the first register (fig.\ref{qcircuit of parity}) is assigned to  
flourine (qubit) and the second to protons (qutrit). The algorithm starts with a mixed state (ensemble of 6 basis states of a 
qubit-qutrit system), a pseudo
Hadamard gate on the qubit register, creates single quantum coherences, whose phases are modulated by the oracle. Detection is done on the qubit register, which measures the single quantum coherences. 
After the pseudo Hadamard gate, the polarization of the qubit register consists the sum of three single quantum coherences, which can be written as,

\begin{eqnarray}
I_x^{qubit}=\vert 0 \rangle \langle 1 \vert \vert 0 \rangle \langle 0 \vert  + \vert 0 \rangle \langle 1 \vert \vert 1 \rangle \langle 1 \vert + 
\vert 0 \rangle \langle 1 \vert \vert 2 \rangle \langle 2 \vert,                   \label{polarization of qubit}
\end{eqnarray}
where the first part of each term indicates the single quantum coherence of the qubit  and second part indicates the state of qutrit.

The oracle in (fig.\ref{qcircuit of parity}) is defined as 

\begin{eqnarray}
O=\pmatrix{ (-1)^{x_1}&0&0&0&0&0 \cr 0&(-1)^{x_2}&0&0&0&0 \cr 0&0&(-1)^{x_3}&0&0&0 \cr 0&0&0&(-1)^{x_4}&0&0 \cr 0&0&0&0&(-1)^{x_5}&0 
\cr 0&0&0&0&0&(-1)^{x_6}}, \label{O}
\end{eqnarray}

Table(2) shows the unitary operators of the oracle corresponding to different strings, X=$\{x_1,...,x_6\}$, of various parities. 
Here, one should note that the oracles, for parities 0, 1 and 2 are identical to that of parities 6, 5 and 4 respectively, up to an overall phase factor 
$e^{i\pi}(=-1)$. 
Hence we implement only the oracles of parities 0, 1, 2 and 3. 
For even parity, 
even number of terms of eqn.(\ref{polarization of qubit}) acquire a phase factor $e^{i\pi}$, and for odd parity odd 
number of terms acquire a phase factor $e^{i\pi}$. Hence, by measuring single quantum coherences, one can determine the parity that is encoded to 
the 6 digit binary string.

\begin{center}
\begin{figure}
\epsfig{file=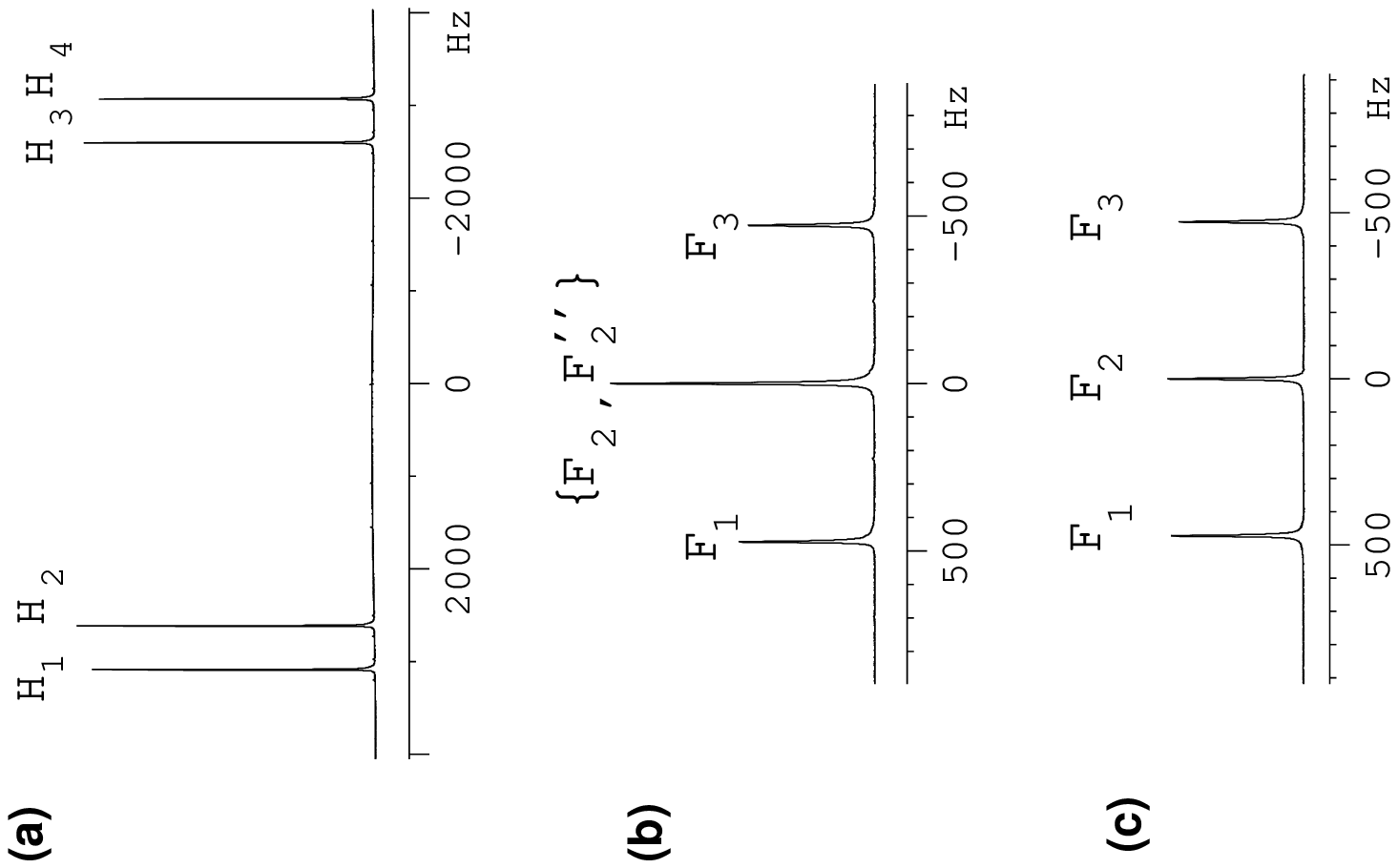,width=12cm,angle=270}
\caption{}    \label{equilibrium spectra of CH2FCN}
\end{figure}
\end{center}

The equilibrium spectra of $^{1}H$ and $^{19}F$ are shown in fig.\ref{equilibrium spectra of CH2FCN}a and 
\ref{equilibrium spectra of CH2FCN}b respectively. In the flourine 
spectrum, the intensity of 
central transition is twice that of satellite transitions, due to the overlap of the transition of the asymmetric manifold.

\begin{center}
\begin{figure}
\epsfig{file=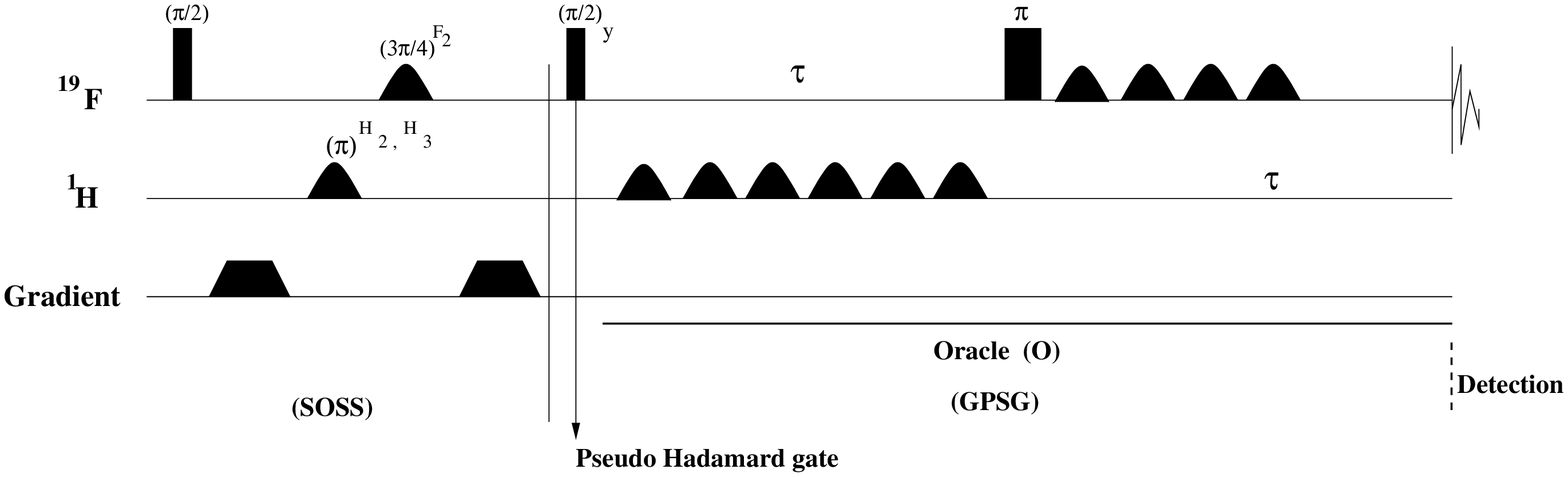,width=16cm}
\caption{}    \label{pulse sequence for parity}
\end{figure}
\end{center}

The experiment consists of 3 steps (as shown in fig.\ref{pulse sequence for parity})

(i) Selection of symmetric manifold (SOSS):
The magnetization from asymmetric manifold is 
saturated by using the pulse sequence (fig.\ref{pulse sequence for parity}), known as SOSS. A gradient pulse following 
a $\pi/2$ pulse on flourine, saturates the flourine 
magnetization. The flourine magnetization of symmetric manifold is retrieved by partial polarization transfer from $^{1}H$ to $^{19}F$ by applying 
selective $\pi$ pulses on $H_2$ and $H_3$ transitions followed by a selective $3\pi/4$ pulse on $F_2$ transition. A final gradient pulse dephases the 
transverse magnetization created during the process.   
The spectrum obtained by a small angle pulse after SOSS, gives the relative intensities in the expected ratio 1:1:1 
(fig.\ref{equilibrium spectra of CH2FCN}c), confirming SOSS. The six energy 
levels of the symmetric manifold can be treated as a 6 digit binary string X=$\{x_1,...,x_6\}$. 

(ii) Implementation of pseudo Hadamard gate followed by oracle:
A $(\pi/2)_y$ on resonance pulse, applied on $^{19}F$, implements a pseudo Hadamard gate, and the created coherences does not evolve under the 
chemical shift, since the pulse is applied at resonance.  
The unitary operators of the oracle, are implemented by using non-adiabatic geometric phases. The phase shift gates 
$d^{\phi}_{H_i}$ or $d^{\phi}_{F_i}$ (Table 3) can be implemented by applying two transition selective $\pi$ pulses respectively on transitions $H_i$ or 
$F_i$ with phases $y$ and $(y+\pi+\phi)$, and the duration of the pulses is set such that the evolution under the hetero nuclear couplings 
$(J+2D)$ makes a complete $2\pi$ rotation. However spin echo sequence ($\tau-\pi-\tau$) is also used which refocuses hetero nuclear couplings and  
chemical shift evolution (if any). Since the unitary operators of the oracle are diagonal matrices, one can sandwich different phase shift gates 
to get a desired phase shift gate, as shown in Table(2), up to an overall phase factor. For example $O^{4}_1$ is a
product of $d^{\pi/3}_{H_2}$, $d^{\pi/6}_{H_3}$, $d^{-\pi/2}_{H_4}$, $d^{\pi/6}_{F_1}$, $d^{\pi/3}_{F_3}$ (up to an overall phase factor $e^{i\pi/6}$), 
which is implemented by three pairs of selective $\pi$ pulses on $^{1}H$ channel and two pairs of $\pi$ pulses on $^{19}F$ channel, fig.16.  Similarly 
one can implement remaining oracles by sandwiching various phase shift gates, as shown in Table 2.

\begin{center}
\begin{figure}
\epsfig{file=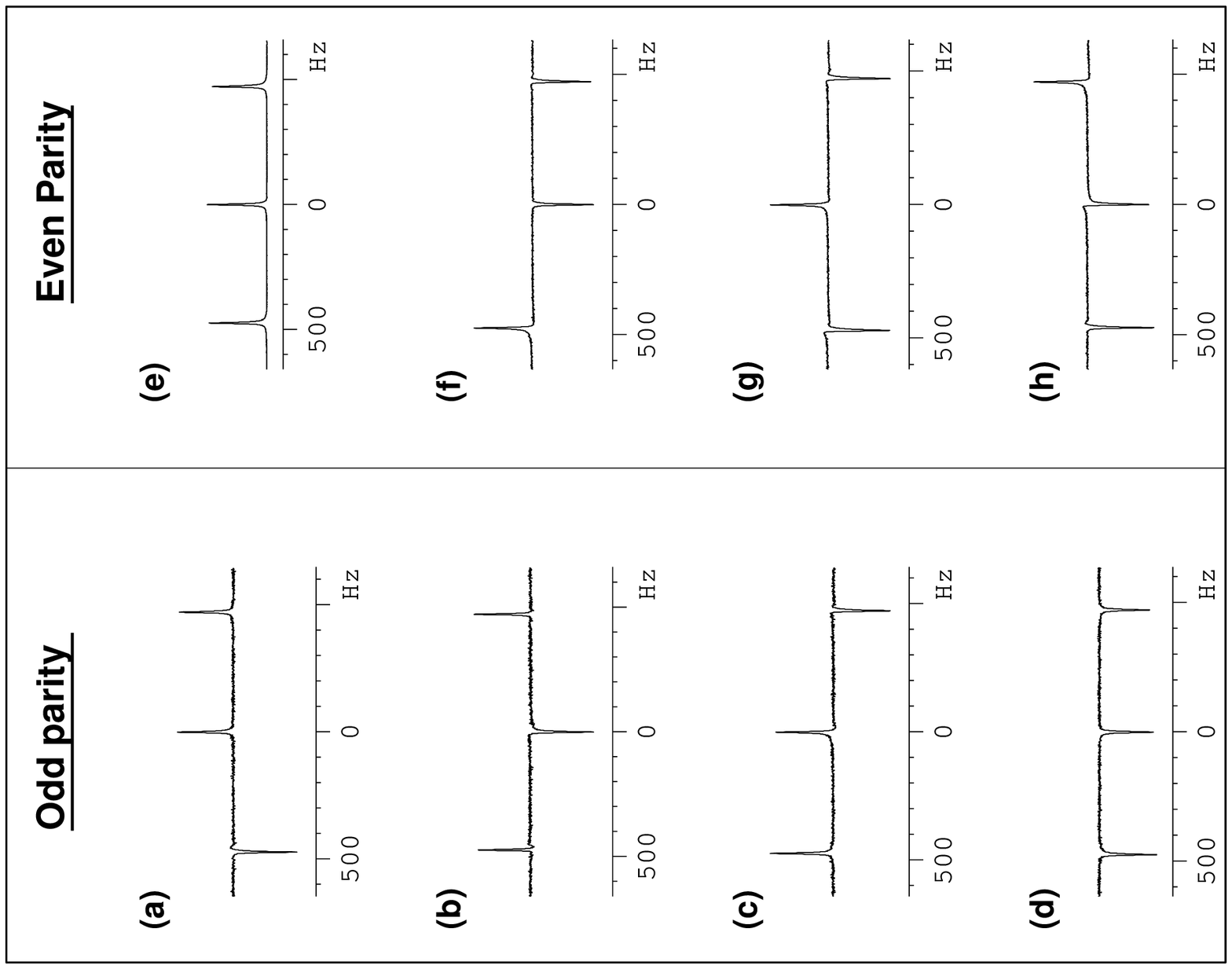,width=12cm,angle=270}
\caption{}    \label{result of parity}
\end{figure}
\end{center}

(iii) Detection: Single quantum coherences of flourine are measured. Depending on the number of 
peaks that are inverted the parity is determined such that, for even parity, even number (0 or 2) of peaks are inverted, 
while for odd parity, odd number (1 or 3) of peaks are inverted. The figures \ref{result of parity}(a), \ref{result of parity}(b), 
\ref{result of parity}(c) and \ref{result of parity}(d) indicate the odd parity, which are obtained by 
implementing $O^{2}_1$, $O^{4}_1$,  
$O^{6}_1$ and $O^{(2,4,6)}_3$ (Table 2) respectively. Figures \ref{result of parity}(e), \ref{result of parity}(f), \ref{result of parity}(g) and 
\ref{result of parity}(h) indicate the even parity, which are obtained by implementing I, 
$O^{(4,6)}_2$, $O^{(2,6)}_2$ and $O^{(2,4)}_2$ (Table 2) respectively. These spectra clearly confirm the implementation of the parity algorithm in this 
qubit-qutrit system, utilizing non-adiabatic geometric phase gates.
  
\section{VI. Conclusions}
Conventionally, controlled phase shift gates which are basic elements in many quantum circuits, are implemented by the 
combination of  qubit selective pulses and evolution under the weakly coupled Hamiltonian. 
In order to achieve higher number of qubits in NMR, one explores dipolar couplings which are larger in magnitude, yielding strongly coupled spectra. 
In such systems since the Hamiltonian consists of non-secular terms it is difficult to apply qubit selective pulses. The present method to implement  
controlled phase shift gates, does not require qubit selective pulses and evolution under Hamiltonian. While in this paper, we have used the systems 
containing only one homo nuclear dipolar coupling, the method is being extended to systems containing more than one homo nuclear dipolar coupling, by 
using suitable refocusing schemes and by using strongly modulated pulses \cite{cory}. This method can also be applied to quadrapolar systems. 
This method thus contributes towards the use of strongly coupled spins for NMR Quantum computing.

\section{ACKNOWLEDGMENTS}

Useful discussions with Dr. Ranabir Das and H. S. Vinay Deepak are gratefully acknowledged. The use of AV-500 NMR spectrometer funded by the department of 
science 
and Technology (DST), New Delhi, at the NMR research center (former Sophisticated Instruments Facility), Indian Institute of Science, 
Bangalore, is gratefully acknowledged. A.K. acknowledges DAE for Raja Ramana Fellowship, and DST for a research grant on 
"Quantum Computing using NMR techniques".

\pagebreak

\begin{center}
 Table 1. Constant and Balanced functions for a two qubit system \\
\end{center}

\begin{center}
\begin{tabular}
{|c|c|c|c|c|c|c|c|c|} \hline \hline
   &\multicolumn{2}{c|} {Constant}&\multicolumn{6}{c|}{Balanced}\\ \hline \hline
x  & $f_1(x)$  &  $f_2(x)$ & $f_3(x)$ & $f_4(x)$& $f_5(x)$ & $f_6(x)$ & $f_7(x)$ & $f_8(x)$ \\ \hline 
00 & $0$       & $1$       &  $0$     & $1$     & $0$      & $1$      & $0$      & $1$        \\ \hline
01 & $0$       & $1$       &  $1$     & $0$     & $0$      & $1$      & $1$      & $0$        \\ \hline
10 & $0$       & $1$       &  $1$     & $0$     & $1$      & $0$      & $0$      & $1$        \\ \hline
11 & $0$       & $1$       &  $0$     & $1$     & $1$      & $0$      & $1$      & $0$       \\

\hline 
\end{tabular}
\end{center}
\vspace{1cm}

\begin{center}
 Table 2. Unitary operators of the oracle, for various parities of a binary string X \\
\end{center}

\begin{center}
\begin{tabular}
{|c|c|c|} \hline \hline
   Parity & X & O (oracle) \\ \hline \hline
0  & $\{0,0,0,0,0,0\}$  &  diag[1,1,1,1,1,1]=I  \\ \hline
1  & $\{0,1,0,0,0,0\}$  &  $O_1^2=diag[1,-1,1,1,1,1]=d_{H_2}^{-2\pi/3}.d_{H_3}^{\pi/6}.d_{H_4}^{-\pi/2}.d_{F_1}^{\pi/6}.d_{F_3}^{\pi/3}$  \\ 
1  & $\{0,0,0,1,0,0\}$  &  $O_1^4=diag[1,1,1,-1,1,1]=d_{H_2}^{\pi/3}.d_{H_3}^{\pi/6}.d_{H_4}^{-\pi/2}.d_{F_1}^{\pi/6}.d_{F_3}^{\pi/3}$ \\ 
1  & $\{0,0,0,0,0,1\}$  &  $O_1^6=diag[1,1,1,1,1,-1]=d_{H_2}^{\pi/3}.d_{H_3}^{\pi/6}.d_{H_4}^{\pi/2}.d_{F_1}^{\pi/6}.d_{F_3}^{\pi/3}$\\ \hline
2  & $\{0,0,0,1,0,1\}$  &  $O_2^{(4,6)}=diag[1,1,1,-1,1,-1]=d_{H_4}^{\pi}$  \\ 
2  & $\{0,1,0,0,0,1\}$  &  $O_2^{(2,6)}=diag[1,-1,1,1,1,-1]=d_{H_2}^{\pi}.d_{H_4}^{\pi}$ \\ 
2  & $\{0,1,0,1,0,0\}$  &  $O_2^{(2,4)}=diag[1,-1,1,-1,1,1]=d_{H_2}^{\pi}$   \\ \hline
3  & $\{0,1,0,1,0,1\}$  &  $O_3^{(2,4,6)}=diag[1,-1,1,-1,1,-1]=d_{H_3}^{\pi/2}.d_{H_4}^{-\pi/2}.d_{F_1}^{\pi/2}.d_{F_3}^{\pi}$  \\ 

\hline
\end{tabular}
\end{center}

\vspace{1cm}

\begin{center}
 Table 3. Unitary operators of controlled-$\phi$ phase shifted gates \\
\end{center}

\begin{center}
\begin{tabular}
{|c|} \hline
 $d_{H_i(F_i)}^\theta$ \\ \hline
 $d_{F_1}^\phi=diag[e^{i\phi},e^{-i\phi},1,1,1,1]$ \\
 $d_{F_2}^\phi=diag[1,1,e^{i\phi},e^{-i\phi},1,1]$ \\
 $d_{F_3}^\phi=diag[1,1,1,1,e^{i\phi},e^{-i\phi}]$  \\
 $d_{H_1}^\phi=diag[e^{i\phi},1,e^{-i\phi},1,1,1]$  \\
 $d_{H_2}^\phi=diag[1,e^{i\phi},1,e^{-i\phi},1,1]$  \\
 $d_{H_3}^\phi=diag[1,1,e^{i\phi},1,e^{-i\phi},1]$  \\
 $d_{H_4}^\phi=diag[1,1,1,e^{i\phi},1,e^{-i\phi}]$  \\
\hline
\end{tabular}
\end{center}

\pagebreak

\pagebreak
\section{Figure captions}
(1). (a) Schematic energy level diagram of a four level system, such that the single quantum magnetization of the subspaces (1,2), (2,3) and (3,4) is directly 
observable.
(b) Evolution of state $\vert \psi_{rs} \rangle$ (r,s=1,2 and 3, s=r+1), over a closed loop on the Bloch sphere, by applying two transition selective 
$\pi$ pulses on transition (r,s) 
with phases $y$ and $(y+\pi+\pi=y)$. Solid angle subtended by this closed loop at the center of the sphere is, $2\pi$.

(2). Rotations in subspaces spanned by orthogonal operators, of the 4 level system of fig.(1a). (a), (b), (c) and (d) follow 
the commutation relations, respectively given by, $[I_x^{(1,2)}, I_x^{(1,3)}]=\frac{i}{2}I_y^{(2,3)}$, $[I_y^{(1,2)}, I_x^{(1,3)}]=\frac{i}{2}I_x^{(2,3)}$, $[I_x^{(2,4)}, I_y^{(3,4)}]=\frac{i}{2}I_x^{(2,3)}$ and $[I_x^{(2,4)}, I_x^{(3,4)}]=\frac{i}{2}I_y^{(2,3)}$ (eqn.10). Each of the commutation rules follow 
the transformation property given in eqn.(11), for example, for fig.(a) $[I^{(1,2)}_x,I^{(1,3)}_x]=\frac{i}{2}I^{(2,3)}_y$, means that 
a $(\phi)y$ pulse applied on transition (2,3), acts as a $\phi/2$ pulse on transition (1,2), hence $I^{(1,2)}_x$ is 
transformed to $(I^{(1,2)}_x cos(\phi/2)+I^{(1,3)}_x sin(\phi/2))$. (Adopted from ref.s [39], [40])

(3). The schematic energy level diagram of dipolar coupled protons and carbon of $^{13}CH_3I$, partially oriented in a liquid crystal. There are 16 
energy levels, eight of which belong to a symmetric manifold and the other eight to two groups of asymmetric manifolds. 
$h_i$, $h'_i$ and $h''_i$ represent proton transitions and $C_i$, $C'_i$ and $C''_i$ represent carbon transitions. 
All transitions with identical subscript are degenerate, eg. $C_i$, $C'_i$ and $C''_i$ have identical frequencies, as well as $h_i$, $h'_i$ and $h''_i$, 
(see fig.4).

(4). Proton (a) and carbon (b) equilibrium spectra of oriented $^{13}CH_3I$ recorded at 500 MHz of proton frequency. The labeling of transitions are in 
accordance with the notation used in fig.3. 
The transitions $h_2$, $h_4$ and $h_6$ (and also $h_1$, $h_3$ and $h_5$) are split by homonuclear dipolar coupling, and the splitting is
equal to $3D_{HH}$ which has a value 3553 Hz, fig. (4a). The C-H splitting is equal to $(2D_{CH}+J_{CH})$, which has a value 2053 Hz, fig. (4b).
The relative experimental integrated intensities are in the ratio; for $^{1}H$ 
0.98:1.00:2.00:2.02:1.00:0.99 (theoretical ratios are 1:1:2:2:1:1), and for $^{13}C$ 0.98:3.02:3.00:1.00 (theoretical ratios are 1:3:3:1).  
(c) $^{1}H$ spectrum, obtained using a small angle measuring pulse, after implementation of SOSS. The SOSS is implemented here by the pulse sequence 
$(\pi)^{h_3,h_4}-G_z-(\pi)^{h_3,h_4}$. 
The relative experimental integrated intensities are in the ratio 2.97:3.02:4.05:4.02:3.00:2.98 (theoretical ratios are 3:3:4:4:3:3).

(5). The pulse sequence for implementation of Geometric phase shift gate (GPSG) U(1,2)=diag[$e^{i\phi}$,$e^{-i\phi}$,1,1,1,1,1,1]. The pulse sequence 
consists of four parts; 
(i) SOSS:[$(\pi)^{h_3,h_4}-G_z-(\pi)^{h_3,h_4}$], (ii) Pseudo Hadamard gate on protons:  $(\pi/2)_y^{^{1}H}$, (iii) Geometric phase shift gate: 
[$\tau-\pi-\tau$], ($(\pi)_y^{C_1}$ $(\pi)_{y+\pi+\phi}^{C_1}$) during second $\tau$ period, evolves the z magnetization of (1,2) subspace 
over a closed loop on the Bloch sphere, with solid angle $2\phi$ (iv) Detection of proton resonances. 

(6). Proton spectra obtained after the implementation of Geometric phase shift gate U(1,2)=diag[$e^{i\phi}$,$e^{-i\phi}$,1,1,1,1,1,1] 
(pulse sequence of fig.5). 
(a) to (d) respectively, are for $\phi=\pi/2, \pi, 3\pi/2$ and $2\pi$. Since the Hamiltonian evolution during transition selective pulses is refocused, 
the phases of the the peaks in the spectra are entirely due to geometric phases and confirm the implementation of geometric phase 
shift gates. The transition selective carbon pulses are low power Gaussian pulses of duration 5.65 msec (=$\tau/2$).     

(7). Experimental proton spectra obtained after implementation of $(\pi)$-phase shifted gates on two unconnected subspaces. These gates are obtained by 
applying two transition selective $(2\pi)_y$ pulses on carbon transitions, during the second $\tau$ period in the pulse sequence of fig.5. 
The length of transition selective $2\pi$ pulse is identical as that of $\pi$ pulse but has double the pulse power, such that the dynamic phase remains 
refocused.
Spectra corresponding to 
geometric phase shift gates (a) U[(3,4),(5,6)]=diag[1,1,-1,-1,-1,-1,1,1] implemented by applying $(2\pi)_y$ pulses on transitions $C_2$ and $C_3$. In this 
experiment the coherences $h_1$, $h_2$, $h_5$ and $h_6$ acquire a Geometric phase of $\pi$ while 
the coherences $h_3$ and $h_4$ acquire a phase of $2\pi$. 
(b) U[(1,2),(3,4)]=diag[-1,-1,-1,-1,1,1,1,1] implemented by applying $(2\pi)_y$ pulses on 
transitions $C_1$ and $C_2$ yielding the proton spectrum with coherences $h_3$ and $h_4$ yielding a $\pi$ phase shift.
(c) U[(1,2),(5,6)]=diag[-1,-1,1,1,-1,-1,1,1], implemented by applying $(2\pi)_y$ pulses on transitions $C_1$ and $C_3$, yielding the proton spectrum, 
in which all proton coherences acquire $\pi$ phase shift.

(8). Quantum circuit for implementation of two qubit Deutsch-Jozsa algorithm (Collins version). The pseudo Hadamard gate creates 
superposition of the four basis states. $U_{fi}$ is the unitary transformation, given in eqn.14, corresponding to
 the function $f_i$ given in table(I). The final Hadamard gate in the original circuit is canceled by a detection pulse (pseudo Hadamard gate). Hence the result of the algorithm is available immediately after $U_{fi}$.
  
(9). Schematic energy level diagram of dipolar coupled oriented $CH_3CN$. 
The four levels of 
symmetric manifold are labeled as, basis states of a two qubit system. Transitions 2, 4 and 5 are degenerate, as shown in fig. 10(a).

(10). (a) Equilibrium proton spectrum of oriented $CH_3CN$. The labeling of transitions are in accordance with the notations used in fig.(9). 
The relative integrated intensities are in the ratio 1.00:1.99:0.98 (theoretical ratios are 1:2:1), and the splitting $3D_{HH}$ has a value 4968 Hz.   
(b) spectrum obtained by a small angle pulse after SOSS by the pulse sequence $(\pi)^2-G_z-(\pi)^2$ (fig.11). The relative experimental integrated 
intensities are in the 
ratio 3.02:3.98:2.97 (theoretically expected 3:4:3). 
(c) spectrum obtained by a small angle pulse after the preparation of pseudo pure state $\vert00\rangle$, by using the pulse sequence, 
$(\pi)^3-(\pi/2)^2-G_z$ (fig.11).

(11). The pulse sequence for the implementation of two qubit Deutsch-Jozsa algorithm (fig.8), 
consists of four parts; (i) SOSS: $(\pi)^2-G_z-(\pi)^2$, (ii) Pseudo pure state $\vert00\rangle$: $(\pi)^3-(\pi/2)^2-G_z$,
(iii) Coherent superposition followed by $U_{f_7}$=diag[1 -1 1 -1]: A $(\pi/2)_y$ on resonance pulse creates a superposition state 
$\vert \psi\rangle=(1/2\sqrt2)(\vert00\rangle+\sqrt3\vert01\rangle+\sqrt3\vert10\rangle+\vert11\rangle)$, $U_{f_7}$ is implemented by 
[$(2\pi)_y^2$ $(2\pi)_{y}^3$]. The duration of transition selective pulses (5.229 msec) is such that the evolution due to  
dipolar coupling makes a complete $2\pi$ rotation and thus the phases are only due to cyclic geometric evolution of subspaces. (iv) Detection of single quantum 
coherences. 

(12). Implementation of two qubit Deutsch-Jozsa algorithm on oriented $CH_3CN$, by using the pulse sequence shown in fig.11. Transition selective pulses are of Gaussian shape with duration 5.229 msec. After pseudo Hadamard gate (fig.11), controlled phase 
shift gate $U_{f_i}$ is implemented. 
(a) $U_{f_1}$ is identity matrix which require no pulse, hence single quantum coherences are measured immediately after pseudo Hadamard gate. The 
integrated intensities are in the ratio, 1.69:3.00:1.72 (theoretically expected $\sqrt3$:3:$\sqrt3$ \cite{mahesh}), confirms state $\vert \psi_{f1} \rangle$ of eqn.15.
(b) $U_{f_3}$ (c) $U_{f_5}$, are implemented by applying a single $(2\pi)_y$ pulse on transitions 2 and 3 respectively. 
(d) $U_{f_7}$ is implemented by applying two $(2\pi)_y$ pulses on transitions 2 and 3.
In fig.(a), none of the peaks are inverted, indicates the constant function. The fig.s(b,c and d) confirms the final states $\vert \psi_{f3} \rangle$, 
$\vert \psi_{f5} \rangle$ and $\vert \psi_{f7} \rangle$ of eqn.(15), and atleast one of the peaks are inverted, indicates 
the balanced functions.

(13). Quantum circuit for implementing qubit-qutrit parity algorithm. Initial state is a mixed state followed by a pseudo Hadamard gate (h) on qubit register. 
Unitary operators of oracle are controlled phase shift gates, given in table(2) for various parities. The final step is the measurement 
on qubit register, which in NMR 
is the direct signal acquisition of resulting single quantum coherences.

(14). Schematic energy level diagram of oriented $CH_2FCN$, consisting of symmetric and asymmetric manifolds. Flourine transitions are labeled as $F_i$ and $F'_i$ 
and that of protons as $H_i$. The transitions $F_2$ and $F'_2$ are degenerate (see fig.15b).

(15). (a) and (b) are, respectively, equilibrium proton and flourine spectra of oriented $CH_2FCN$. 
In the proton spectrum, experimental integrated intensities are in the ratio 0.98:1.00:0.99:0.98 (theoretical ratios are 1:1:1:1), and that of 
flourine are in the ratio 0.99:2.01:1.00 (theoretical ratios are 1:2:1).  
The transitions $H_1$ and $H_3$
(also $H_2$ and $H_4$) are split by homonuclear diploar coupling. The splitting is equal to $3D_{HH}$ which has a value 5694 Hz. The 
$^{19}F$-$^{1}H$ 
splitting in (b) is equal to
($2D_{FH}$+$J_{FH}$), which has a value 473 Hz.
(c) Flourine spectrum obtained by a small angle pulse after SOSS by the pulse sequence $\pi/2-G_z-(\pi)^{H_2,H_3}-(3\pi/4)^{F_2}-G_z$ (fig.16). The 
relative experimental integrated intensities are in the ratio 1.00:1.02:0.98 (theoretical ratios are 1:1:1).

(16). The pulse sequence to implement qubit-qutrit parity algorithm (fig.13). 
The pulse sequence consists of four parts; (i) SOSS: $\pi/2-G_z-(\pi)^{H_2,H_3}-(3\pi/4)^{F_2}-G_z$, (ii) Pseudo Hadamard gate: On resonance 
$(\pi/2)_y^{^{19}F}$ pulse, 
(iii) Oracle $O_1^4$, shown in Table 2:  During first $\tau$ period three pairs of transition selective proton $\pi$ pulses [$(\pi)_y^{H_2}$$(\pi)_{(y+\pi+\pi/3)}^{H_2}$]
[$(\pi)_y^{H_3}$$(\pi)_{(y+\pi+\pi/6)}^{H_3}$]
[$(\pi)_y^{H_4}$$(\pi)_{(y+\pi-\pi/2)}^{H_4}$] are applied, which respectively implement $d_{H_2}^{\pi/3}$, $d_{H_3}^{\pi/6}$ and $d_{H_4}^{-\pi/2}$, 
and during second $\tau$ period two pairs of transition selective flourine $\pi$ pulses 
[$(\pi)_y^{F_1}$$(\pi)_{(y+\pi+\pi/6)}^{F_1}$][$(\pi)_y^{F_3}$$(\pi)_{(y+\pi+\pi/3)}^{F_3}$] are applied, which respectively implement $d_{F_1}^{\pi/6}$ 
and $d_{F_3}^{\pi/3}$. The duration of transition selective $\pi$ pulses is set such that the evolution under the hetero nuclear couplings $(J+2D)$ makes 
a complete $2\pi$ rotation.  
(iv) Detection of single quantum coherences of flourine, from which the parity can be determined.

(17). Implementation of qubit-qutrit parity algorithm, by using the pulse sequence shown in fig.16. 
The transition selective $\pi$ pulses, for both proton and flourine, are of Gaussian shape with duration 12.85 msec.  
For odd (even) parity odd (even) number of peaks are inverted, in other words odd (even) number of single quantum coherences of flourine 
acquire a phase $\pi$. (a), (b), (c) and (d) indicate the odd parity which are obtained by implementing 
$O^{2}_1$, $O^{4}_1$, $O^{6}_1$ and $O^{(2,4,6)}_3$ (Table 2) respectively. (e), (f), (g) and (h) indicate the even parity, which are 
obtained by implementing I, $O^{(4,6)}_2$, $O^{(2,6)}_2$ and $O^{(2,4)}_2$ (Table 2) 
respectively.

\newpage

\pagebreak

\pagebreak

\pagebreak

\pagebreak

\pagebreak

\pagebreak

\pagebreak

\pagebreak

\pagebreak

\pagebreak

\pagebreak

\pagebreak

\pagebreak

\pagebreak

\pagebreak

\pagebreak

\end{document}